\documentclass[a4paper,12pt]{article}

\usepackage[height=8.85in,width=6.45in]{geometry}
\usepackage{fancybox}
\usepackage{amsmath,amssymb}
\usepackage[T1]{fontenc}
\usepackage{comment}
\usepackage{hyperref}
\usepackage{xcolor}

\usepackage[all]{xy}

\DeclareMathAlphabet{\mathitsf}{\encodingdefault}{\sfdefault}{m}{sl}

\def\a{\alpha}
\def\b{\beta}

\def\aa{{\dot \a}}
\def\bb{{\dot \b}}

\def\CA{{\cal A}}

\def\CE{{\cal E}}

\def\CL{{\cal L}}

\def\CN{{\cal N}}
\def\CO{{\cal O}}

\DeclareMathOperator{\Tr}{Tr}
\DeclareMathOperator{\Pf}{Pf}
\usepackage{slashed}
\let\Slash\slashed

\def\diag{\mathop{\rm diag}\nolimits}
\def\Spin{\mathop{\rm Spin}}
\def\Pin{\mathop{\rm Pin}}
\def\pin{\mathop{\rm pin}}
\def\SO{\mathop{\rm SO}}
\def\OO{{\rm O}}
\def\SU{\mathop{\rm SU}}
\def\U{\mathrm{U}}
\def\Sp{\mathop{\rm Sp}}

\def\SL{\mathop{\rm SL}}
\def\tr{\mathop{\rm tr}}

\def\CA{{\cal A}}

\def\CE{{\cal E}}

\def\CL{{\cal L}}

\def\CN{{\cal N}}
\def\CO{{\cal O}}

\newcommand{\bH}{\mathbb{H}}
\newcommand{\bC}{\mathbb{C}}
\newcommand{\bR}{\mathbb{R}}
\newcommand{\bZ}{\mathbb{Z}}
\def\Nequals#1{$\mathcal{N}{=}#1$}
\def\SU{\mathrm{SU}}
\def\Sp{\mathrm{Sp}}

\def\U{\mathrm{U}}
\def\SO{\mathrm{SO}}

\def\SL{\mathrm{SL}}
\def\tr{\mathop{\mathrm{tr}}\nolimits}
\def\diag{\mathop{\mathrm{diag}}\nolimits}

\def\sign{\mathop{\mathrm{sign}}\nolimits}
\def\Tr{\mathop{\mathrm{Tr}}\nolimits}

\def\RP{\mathbb{RP}}

\def\sT{\mathsf{T}}
\def\sC{\mathsf{C}}
\def\sP{\mathsf{P}}
\def\sCP{\mathsf{CP}}
\def\sCR{\mathsf{CR}}
\def\sCPT{\mathsf{CPT}}

\def\sG{\mathsf{G}}
\def\sF{\mathsf{F}}

\def\sPinv{\sP_\text{inv}}

\def\beq#1\eeq{\begin{align}#1\end{align}}

\usepackage{framed}
\definecolor{shadecolor}{rgb}{0.90,0.90,0.90}
\newenvironment{claim}{\begin{shaded}\noindent\itshape\ignorespaces}{\end{shaded}}

\usepackage{times}
\usepackage{courier}
\usepackage{mathtools}
\numberwithin{equation}{section}

\let\bar\overline
\def\ii{\mathrm{i}\,}
\def\Nequals#1{$\mathcal{N}{=}\,#1$}

\def\dualPhi{{{\mathsf{\Phi}}}}
\def\dualPsi{{{\mathsf{\Psi}}}}
\def\dualm{\mathitsf{m}}
\def\dualg{\mathitsf{g}}

\def\dualQ{\mathitsf{Q}}

\def\dualSU{\mathitsf{SU}}
\def\dualU{\mathitsf{U}}
\def\dualSO{\mathitsf{SO}}
\def\dualE{\mathitsf{E}}
\def\dualB{\mathitsf{B}}
\def\dualh{\mathitsf{h}}
\def\duala{\mathitsf{a}}
\def\underQ{S}

\def\underdualQ{\mathitsf{S}}
\def\frakq{\mathfrak{q}}

\begin{document} 

\begin{titlepage}

\begin{flushright}
IPMU-16-0034
\end{flushright}

\vfill

\begin{center}

{\Large\bfseries Gauge interactions and topological phases of matter}

\vskip 1cm
Yuji Tachikawa and Kazuya Yonekura
\vskip 1cm

\begin{tabular}{ll}
  & Kavli Institute for the Physics and Mathematics of the Universe, \\
& University of Tokyo,  Kashiwa, Chiba 277-8583, Japan
\end{tabular}

\vskip 1.5cm

\textbf{Abstract}

\end{center}

\vskip1cm

\noindent
We initiate the study of the effects of  strongly-coupled gauge interactions on the properties of the topological phases of matter. In particular, we discuss fermionic systems with three spatial dimensions, protected by time reversal symmetry.
We first derive a sufficient condition for the introduction of a dynamical Yang-Mills field to preserve the topological phase of matter,
and then show how the massless pions capture in the infrared the topological properties of the fermions in the ultraviolet.
Finally, we use the S-duality of \Nequals2 supersymmetric $\SU(2)$ gauge theory with $N_f{=}4$  flavors to show that the $\nu{=}16$ phase of Majorana fermions
can be continuously connected to the trivial $\nu{=}0$ phase.

\vfill

\end{titlepage}

\setcounter{tocdepth}{2}
\tableofcontents
\newpage

\section{Introduction}

The main topic of this paper is  the effects of strongly-coupled gauge interactions on topological phases of matter. 
Two general questions immediately come to our mind: \begin{itemize}
\item How would the strongly-coupled gauge interactions affect the topological phases of matter?
\item Can we use the strongly-coupled gauge dynamics  to study the topological phases of matter?
\end{itemize}
In this paper we would like to start providing some answers to these questions.

We begin in Sec.~\ref{1.1} by presenting a motivation to study topological phases of matter from a high-energy physics point of view. We then describe very briefly the classification of free fermionic topological phases in Sec.~\ref{1.2}.
Readers already convinced by the importance of topological phases of matter can safely skip sections~\ref{1.1} and \ref{1.2} to Sec.~\ref{subsec:mainIntroduction}, where we come back to the issue of  gauge interactions. 
After distinguishing and emphasizing the various roles symmetries play in the paper in Sec.~\ref{subsec:remark}, the organization of the rest of the paper is given in Sec.~\ref{subsec:org}.

\subsection{SPT phases and classification of QFTs}\label{1.1}
There are by now bewildering varieties of quantum field theories (QFTs) realized experimentally and/or constructed theoretically.
One way to put them in order is to try to classify them. 
To start a classification, we need to decide which kind of QFTs we treat,
and what equivalence relation we use.
The classification becomes the more tractable when we treat the simpler QFTs under the coarser equivalence relation.

To have a simple classification, let us only treat the ones whose excitations are all gapped or equivalently massive. 
Then, the infrared limit is almost empty, in that there can only be a finite number of vacuum states on a given space.
Let us further demand that there is in fact only a unique vacuum state, whatever the topology of the space is as long as it is compact without any boundary. 
Such theories are said to have no intrinsic topological order. 

We now fix a symmetry group $\sG$ and the spacetime dimension $d+1$, and then  consider all possible $d+1$ dimensional gapped QFTs with $\sG$ symmetry without intrinsic topological order.
Here, the symmetry $\sG$ can be arbitrarily chosen to your liking: it can include a spacetime discrete symmetry such as the time reversal $\sT$, and also an internal continuous symmetry such as $\SU(2)$.
We also need to specify whether we consider bosonic or fermionic QFTs, in the sense that the theory under consideration detects the spin structure of the spacetime manifold or not.
To make the classification the most tractable, we put the coarsest equivalence relation among such QFTs.
This is done by declaring that two QFTs are equivalent when they can be continuously deformed to each other without leaving the class of such gapped QFTs with symmetry $\sG$.
An equivalence class is then called a symmetry protected topological (SPT) phase protected by $\sG$.
We can always consider a completely trivial theory, such that there is only one state in the Hilbert space with a trivial $\sG$ action and the transition amplitude is always 1 in any situation.
The equivalence class containing this trivial theory is the trivial SPT phase, while the other SPT phases are the topological phases of matter.

In order to see if two QFTs $X_1$ and $X_2$ belong to the same SPT phase or not, it is useful to consider the setup where the region $y<0$ is filled with the system $X_1$ and the region $y>0$ by $X_2$,
with a thin transition region between the two.
We call this transition region as the boundary.
When $X_1$ and $X_2$ belong to the same SPT phase, we can choose the configuration on the total space such that the system is gapped everywhere, without any topological order even at the boundary.
In contrast, when $X_1$ and $X_2$ are distinct SPT phases, something must happen at the boundary of $X_1$ and $X_2$: 
there might be a gapless mode, or a spontaneous symmetry breaking of $\sG$, or a gapped surface topological order at the boundary.

\subsection{Fermionic SPT phases}\label{1.2}
A fermionic SPT phase, when $\sG$ is given by a global $\U(1)$ symmetry together with a few discrete symmetry, is called a topological insulator in the literature.
This terminology is due to the fact that  in a real insulator the excitation is gapped and the electromagnetic  $\U(1)$ symmetry is unbroken.
Note that the $\U(1)$ symmetry in this case is considered as a global symmetry.
Similarly, a fermionic SPT phase, when $\sG$ is just given by a few discrete symmetry, is called a topological superconductor, because in a superconductor the excitation is again gapped and the electromagnetic $\U(1)$ symmetry is broken and not there in the infrared.

In general these systems have complicated interactions.
To make the classification even simpler, it is instructive to start by considering only free massive fermions. 
It was found that the possible choice of discrete symmetries can be summarized in a 10-fold way, and the dependence on the number of spacetime dimensions follows a uniform pattern. 
The result is now known as the periodic table of free fermionic SPT phases \cite{Schnyder:2008tya,Kitaev:2009mg,Ryu:2010zza}.\footnote{
See also e.g., \cite{Ho:2012gz,Ho:2012vr,Hashimoto:2015dla,Hashimoto:2016dtm} for a sample of papers in \texttt{hep-th} on the SPT phases.}

Let us recall one concrete case that is the focus of our paper: the $3{+}1$ dimensional free fermionic SPT phases, protected by the time reversal symmetry $\sT$ such that $\sT^2=(-1)^F$,
where $F$ is the fermion number.
They are topological superconductors, and this choice of the protecting discrete symmetry is known as class DIII. 
As we will consider only relativistic systems, we can freely replace the time reversal symmetry $\sT$ with the CP symmetry $\sCP$, which in this case satisfies $\sCP^2=(-1)^F$.
We will mostly use this latter nomenclature, since this will be more familiar to readers of \texttt{hep-th}.

The basic example is a single Majorana fermion. 
The $\sCP$ invariance forces the coefficient $m$ of the mass term to be real. 
In order for the system to be gapped, we need $m\neq 0$.
So the systems can be classified into two disconnected pieces, those with $m>0$ and those with $m<0$. 
Therefore they represent two distinct SPT phases.\footnote{In continuum QFTs, there is no point in saying which of $m>0$ or $m<0$ is the trivial SPT phase, since this notion depends on the UV regularization. One way is to fix the sign of the mass term of the Pauli-Villars regulator fermion to be positive. Then the $m>0$ case is the trivial phase and the $m<0$ case is the $\nu{=}1$ phase. A more invariant way to state the situation is that the $m>0$ case and the $m<0$ case differ by the $\nu{=}1$ SPT phase.}
One manifestation of the distinctness is when we have a space-dependent mass term:
suppose we have a $y$-dependent mass term $m(y)$ such that 
\begin{equation}
m(y)>0\quad (y>\epsilon), \qquad
m(y)<0\quad (y<-\epsilon)
\end{equation} for a small positive $\epsilon$.
Then there is necessarily one massless $\sCP$-invariant $2+1$ dimensional Majorana fermion at the boundary region $y\sim 0$.

More generally, when we have $N$ Majorana fermions, the coefficient of the mass term is a real symmetric matrix $M_{ij}$.
Let us denote by $\nu(M)$  the number of the negative eigenvalues.
Now, consider again a $y$-dependent mass term. 
Then the generic number of massless $\sCP$-invariant $2+1$ dimensional Majorana fermion in the middle is given by the difference of $\nu(M)$ on $y>0$ and $\nu(M)$ on $y<0$.
From this analysis, we see that the $3{+}1$-dimensional free topological superconductors of class DIII are characterized by an integer $\nu$, i.e.~they are classified by $\bZ$.

Weak perturbation cannot change this classification, but interaction effects when it is strong can and indeed do change this classification. 
The  first example that was found was the $1{+}1$-dimensional fermionic SPT phase of class BDI.
At the level of free fermion, it is characterized by an integer $\nu\in \bZ$ mostly as above.
But with suitable four-fermi interactions added, it has been shown that the $\nu{=}8$ phase and the $\nu{=}0$ phase can be continuously connected while gapped \cite{Fidkowski:2009dba}.  
Stated differently, the classification collapses from $\bZ$ to $\bZ_8$ due to the interaction effects.
We now have ample pieces of evidence \cite{Fidkowski:2013jua,Wang:2014lca,KitaevCollapse,Metlitski:2014xqa,Morimoto:2015lua} that similarly in the $3{+}1$ dimensional fermionic SPT phase protected by the time reversal symmetry $\sT^2=(-1)^F$,  the classification collapses from $\bZ$ to $\bZ_{16}$ once we include the effects of interactions.

There is an ongoing effort to classify interacting bosonic and fermionic SPT phases by finding the right mathematical language to describe them, see e.g.~\cite{Chen:2011pg,Kapustin:2014dxa}. 
These analyses have confirmed the collapse of the free classification due to interactions recalled above, and have shown that there can also be genuinely interacting SPT phases that cannot be continuously connected to free SPT phases. 

\subsection{Gauge interactions and SPT phases}\label{subsec:mainIntroduction}
In this paper, we  consider the effects of dynamical gauge fields on the properties of SPT phases. 
Before going further, it is useful to recall how the gauge fields have been used in their study in the literature.

Firstly, for an SPT phase protected by a symmetry $\sG$, it is extremely convenient to consider coupling the system to background gauge fields for the internal symmetry part of $\sG$, and to nontrivial background metric for the spacetime symmetry part of $\sG$.
In a sense, an SPT phase can be characterized by the response of the system to these background fields.
The $\U(1)$ Maxwell field is often utilized in this manner for SPT phases protected by $\U(1)$, and unoriented background manifolds are used for SPT phases protected by time-reversal symmetry.

Secondly, the topological properties of strongly-coupled confining gauge theories in the infrared have been studied, e.g.~in \cite{GaiottoTalk,Tachikawa:2014mna,Gaiotto:2014kfa}. 
The systems considered there do, however, have intrinsic topological order in the infrared, in the sense that there are multiple degenerate vacua on nontrivial spacetime manifolds, and they correspond to what are usually referred to as symmetry enriched topological (SET) phases in the literature, and not to genuine SPT phases in the narrower technical sense.

In this paper, we initiate the study of the effects of dynamical gauge interactions, which can be strongly coupled, to the SPT phases. 
The main questions we pose are twofold.

\textbf{The first question} is when the introduction of dynamical gauge fields to a given system does not destroy the SPT phase of the original system. 
Suppose an SPT phase $X$ protected by a symmetry $\sG$  to be considered is realized by a QFT with an additional symmetry $H$. 
Then, we can introduce a dynamical gauge field for $H$.
We denote the combined system with dynamical gauge field by $X/H$.
Naively, when the gauge interactions become very strong and confine themselves, it should be possible to integrate them out.
When there is no $\sG\times H$ mixed anomaly, the process of integrating out should not introduce any  interaction that breaks $\sG$, and therefore it should not change the SPT phase.
Our main claim in this regard is the following:
\begin{claim}
Suppose the SPT phase $X$ protected by a symmetry $\sG$  with three spatial dimensions we would like to consider has  an additional symmetry $H$ to which we can couple a dynamical gauge field.  When the  group $H$ is simple, connected and simply-connected and furthermore the effective theta angle of $H$ is zero,  the system $X/H$ after the introduction of the gauge field can be obtained as a continuous deformation of the original system $X$ without closing the gap. In particular, $X$ and $X/H$ are in the same SPT phase protected by $\sG$.
\end{claim}

After making a general argument leading to this claim, 
we perform a detailed check  in the case when the original SPT phase is a system of free Majorana fermions protected by time-reversal symmetry $\sT$ with $\sT^2=(-1)^F$.
We will see that the SPT phase of the ultraviolet fermions is indeed captured by the non-linear sigma model of the massless pions in the infrared for certain flavor numbers of quarks.
These analyses will be performed in Sec.~\ref{sec:pathint} and Sec.~\ref{sec:lowenergy}.

\textbf{The second question} is whether the knowledge of the dynamics of strongly-coupled gauge theories we acquired in the last three decades is useful to shed new light on the properties of standard SPT phases. 
For example, can it be used to show that the interaction effects should collapse the classification of the $3{+}1$-dimensional topological superconductor of type DIII from $\bZ$ to $\bZ_{16}$?
We would like to answer this question in the affirmative. 
We know from the seminal work of Seiberg and Witten \cite{Seiberg:1994rs,Seiberg:1994aj} two decades ago that \Nequals2 supersymmetric $\SU(2)$ gauge theory with $N_f{=}4$ flavors has the S-duality, meaning that when the coupling constant of the $\SU(2)$ gauge group is made extremely large, 
there is a dual description of the same system using a dual $\SU(2)$ gauge group and dual matter contents such that the coupling constant is weak.
A simple counting shows that the hypermultiplets of this system consist of 16 Majorana fermions,
and that the introduction of the $\SU(2)$ gauge field should not change the SPT property of these fermions, according to the criterion we mentioned above.
We will show that this S-duality allows us to connect the $\nu{=}16$ phase continuously to the $\nu{=}0$ phase.
This we will do using the following strategy. First, we add to the free $\nu{=}16$ SPT phase $\SU(2)$ gauge interactions and other fields that do not change the SPT properties so that the system is the \Nequals2 $\SU(2)$ gauge theory with $N_f{=}4$ flavors softly broken to zero supersymmetry.
Second, we increase the gauge coupling constant, and pass to the dual weakly-coupled description.
Third, we add various interaction terms in the dual description to show that it is in the $\nu{=}0$ phase.
We will detail the procedure in Sec.~\ref{sec:super}.

\subsection{Remark on the types of symmetries}\label{subsec:remark}
Before moving on, it would be instructive here to emphasize that there are three different types of symmetries considered in this paper.
The reader is advised to distinguish them to avoid possible confusion.
\begin{itemize}
\item{\bf Global symmetries protecting SPT phase:} For example, a topological superconductor might be protected by $\sT$ and a topological insulator by $\sT \ltimes \U(1) $.
We typically use a letter such as $\sG$ to denote a symmetry protecting the SPT phases. 
In most of the paper, we take it to be just the time-reversal symmetry $\sG=\sT$, with the exception that
in Sec.~\ref{subsec:flavor} we also discuss the case $\sG = \sT \times \sF$
for an additional internal symmetry $\sF$. 
\item {\bf Dynamical gauge symmetries:}  They are associated to dynamical gauge fields living inside the bulk material we are considering, and we integrate over these fields in the path integral. 
 We typically use letters such as $H$ (in this section) or $G$ (in other sections) for dynamical gauge groups.
Notice that all the physical states in the Hilbert space are singlets under the gauge group (on a compact space), and in that sense the gauge symmetry is not a symmetry of physical systems.
\item{\bf Accidental symmetries:} Sometimes, a system we consider happens to have more symmetries than $\sG$. We may call them as accidental symmetries. 
For example, free fermion systems can have much larger symmetry than just $\sT$.
We can easily break them explicitly by introducing some (possibly higher dimensional) operators in the Lagrangian if we do not like them to exist. We typically use a letter such as $F$ to denote them.
\end{itemize}
In particular, we emphasize that
the $\U(1)$ of electromagnetism in the case of topological insulators is {\it not} a dynamical gauge symmetry in our terminology, but should be considered as a part of the global symmetry protecting the SPT phase to which the background non-dynamical electromagnetic field is coupled.

\subsection{Organization of the paper}\label{subsec:org}
In the rest of the paper, we always consider relativistic systems with 3{+}1 spacetime dimensions,
protected by the time-reversal symmetry $\sT$ with $\sT^2=(-1)^F$,
or equivalently by the $\sCP$ symmetry with $\sCP^2=(-1)^F$.
We refer to these systems simply as the SPT phases in this paper.\footnote{
Two justifications of this abuse of the terminology are as follows. 
First, it is simply too tedious to repeat the phrase ``the topological superconductor protected by the time-reversal symmetry $\sT$ with $\sT^2=(-1)^F$.'' 
Second, our discussions in this paper can be generalized to SPT phases other than topological superconductors. See Sec.~\ref{subsec:flavor} for a brief discussion on this point.
}

The rest of the paper is organized as follows. 
In Sec.~\ref{sec:pathint}, we first describe a general argument saying that when the gauge group $G$ is simple, simply-connected and connected, and when the effective theta angle is zero,
then the original  phase $X$ and the system with the gauge field $X/G$ are in the same SPT phase.
 We then check this statement by studying the  effect of nontrivial gauge bundles to the $\eta$ invariant produced by the fermion path integral.

In Sec.~\ref{sec:lowenergy}, we study the low-energy gauge dynamics of non-supersymmetric gauge theories belonging to the class found in the previous section~\ref{sec:pathint} which preserve the SPT properties.
We will find that the $\sigma$-model of the massless pions in the infrared correctly reproduces the $\eta$-invariant of the Majorana fermions in the ultraviolet. 

In Sec.~\ref{sec:super}, we use the S-duality of \Nequals2 supersymmetric $\SU(2)$ gauge theory with $N_f{=}4$ flavors to continuously connect the $\nu{=}16$ phase to the $\nu{=}0$ phase, thus explicitly implementing the collapse of the classification by interaction from $\bZ$ to $\bZ_{16}$.
The basic idea is to note that the hypermultiplets in this \Nequals2 supersymmetric theory consist of 16 Majorana fermions,
and that an extremely strongly coupled region of this theory can be analyzed in a dual weakly-coupled frame.

We conclude the paper with a short discussion in Sec.~\ref{sec:conclusions}.
We have a few appendices: in Appendix~\ref{sec:CPTbasics} the rudimentary facts on $\sCP$ and $\sT$ transformations in $3{+}1$ dimensions are summarized,
paying due attentions to various subtle signs important to us.
Then in Appendix~\ref{sec:concreteCP}, we discuss how the $\sCP$ transformations are implemented in various concrete gauge theories.
We discuss both non-supersymmetric and supersymmetric examples.
Appendix~\ref{sec:WZW} summarizes the properties of Wess-Zumino-Witten terms.
Appendix~\ref{sec:Sd} describes the process of S-duality in $\CN=2$ supersymmetric gauge theory in more detail.
Finally, Appendix~\ref{sec:mass} is a complement to Sec.~\ref{sec:super}.

Before proceeding, we would like to recommend the readers on the \texttt{hep-th} side of the community to go through the excellent paper and lecture notes \cite{Witten:2015aba,Witten:2015aoa} by E. Witten, by which this paper is heavily influenced. 

\section{Effects of gauge fields on SPT phases}\label{sec:pathint}
In this section, we propose and justify a sufficient condition when the coupling of dynamical gauge fields to an SPT phase can be considered as a continuous deformation. 
We first give a general argument in Sec.~\ref{subsec:general},
and provide a detailed analysis verifying the argument when the original SPT phase is given by free fermions in subsections \ref{subsec:preliminary}, \ref{subsec:eta} and \ref{subsec:top}. 
In Sec.~\ref{subsec:flavor} we discuss a simple application of our findings in this section on the structure of the interaction terms that can collapse a free-fermion classification.

\subsection{General construction}\label{subsec:general}
Suppose we are given a system $X$ whose infrared (IR) limit realizes an SPT phase protected by a $\sCP$ symmetry with $\sCP^2=(-1)^F$.
See Appendix~\ref{sec:CPTbasics} for details on the relation between $\sCP$ and $\sT$.
The theory $X$ is by definition gapped. We denote by $M_X$ the mass scale of the gap.\footnote{Throughout the paper we use 
the natural unit of high energy physics where $\hbar=c=1$.}
As can be easily seen from the explicit construction below, the discussion can be generalized to a more general global symmetry $\sG$ protecting the SPT phase. 

Let us further assume that $X$ has an additional continuous non-Abelian $G$ flavor symmetry. Because of the mass gap of $X$, there is no 't Hooft anomaly for this continuous symmetry.
We can then couple a $G$ gauge field to the original system $X$.
We denote the combined system by $X/G$. 

We stress here again that the $G$ here needs to be  distinguished from the $\sG$ discussed in the introduction which is used for the definition of SPT phases.
In this paper we are mainly concerned with the case $\sG=\sCP$ (or equivalently $\sG=\sT$) unless otherwise stated, although many of our results can be generalized to other $\sG$. 
The dynamical gauge group $G$ is, in contrast, a non-abelian Lie group such as $\SU(N)$.

Let us first assume that the dynamical scale $\Lambda_G$ of the gauge theory is far below the gap of the original system $X$, i.e.~$\Lambda_G \ll M_X$.
The Lagrangian of the system in the scale intermediate between $\Lambda_G$ and $M_X$ is given by\footnote{Here we are neglecting possible discrete theta angles.
They do not exist after imposing the condition \eqref{eq:connected-simplyconnected}.} 
\begin{equation}
\CL=- \frac{1}{4g^2_\text{eff}}F^A_{\mu\nu}F^{A \mu\nu}+\frac{\theta_\text{eff}}{64 \pi^2}\epsilon^{\mu\nu\rho\sigma}F^A_{\mu\nu}F^A_{\rho\sigma}.
\end{equation}
This is the effective action of the gauge field which is obtained after integrating out the degrees of freedom of $X$.
We normalize  the theta angle so that one BPST instanton of the gauge group gives 
amplitudes proportional to $e^{\ii \theta}$.

The $\sCP$ invariance of the combined system at this level requires that $\theta_\text{eff}$ is $0$ or $\pi$.
Depending on $\theta_{\rm eff}$, the followings are believed to happen in a pure Yang-Mills theory.
A pure Yang-Mills confines in the IR and has a mass gap. 
If $\theta_{\rm eff}=0$, then there is a unique vacuum which preserves $\sCP$. 
However, if $\theta_{\rm eff}=\pi$, 
it is believed (see e.g.~\cite{Dashen:1970et,Baluni:1978rf,Witten:1980sp,Konishi:1996iz}) that the $\sCP$ is spontaneously broken
and there are two vacua related by $\sCP$. 
The SPT phase classification assumes that the symmetry under consideration is not broken in the bulk.
Therefore, we exclude the case $\theta_{\rm eff}=\pi$ in the following analysis,
and consider only the case when the theta angle is zero:
\begin{equation}
\theta_\text{eff}=0.\label{eq:theta=0}
\end{equation}

In addition, to simplify our analysis, we demand that $G$ is connected and simply-connected: \begin{equation}
\pi_0(G)=\pi_1(G)=0.\label{eq:connected-simplyconnected}
\end{equation}
These are again to keep the system in the standard framework of the SPT phases.
For example, in the gauge group $G$, we could  have included a discrete gauge group such as $\bZ_k $, but such a gauge group gives a topological degrees of freedom in the IR
which contradicts the basic assumption of the SPT phases. Such cases are excluded by the condition $\pi_0(G)=0$.
Even if $G$ does not contain such a discrete gauge group, there is still a possibility that a discrete gauge group appears as a low energy effective theory of confining gauge group.
Let us consider the case of $\SO(3)$ pure Yang-Mills as an example. In this case, the low energy theory contains a $\bZ_2$ gauge group~\cite{Aharony:2013hda,Tachikawa:2014mna}
which can be detected by a $\bZ_2$ 1-form symmetry acting on the 't Hooft loop operators.  
More generally, whenever $G$ is not simply connected, the gauge theory has a 1-form symmetry \cite{Gaiotto:2014kfa} and 
it is believed that we get a nontrivial topological degrees of freedom in the IR. 
Therefore we impose the condition $\pi_1(G)=0$. We will give another but related reason for the condition $\pi_1(G)=0$ below.

Our main claim can now be formulated as given below; the aim of the rest of the paper is to give substance to this claim:
\begin{claim}
When the conditions \eqref{eq:theta=0} and \eqref{eq:connected-simplyconnected} are satisfied, i.e.~when $G$ is (semi)simple, connected and simply-connected and the effective theta angle is zero,  the system $X/G$ after the introduction of the gauge field can be obtained as a continuous deformation of the original system $X$ without closing the gap. In particular, $X$ and $X/G$ are in the same SPT phase.
\end{claim}

Let us first construct a continuous deformation explicitly.
It is generally believed that when there is a field in a gauge theory such that all Wilson lines can be dynamically screened by pair creation of particles, the Higgs phase and the confined phase are continuously connected without any phase boundary. 
This observation goes back to the papers \cite{Banks:1979fi,Fradkin:1978dv}.
More specifically, this folklore theorem stipulates the existence of a  family $Y(\mu)$ of  bosonic systems with flavor symmetry $G$ parameterized by a mass parameter $\mu$, with the following properties.
Namely, when $Y(\mu)$ is considered alone,
\begin{itemize}
\item when $\mu>0$ all the bosons have masses of order $\mu$ and $G$ is unbroken, and
\item  when $\mu<0$ the bosons have vevs of order $|\mu|$ that break $G$ completely,
\end{itemize}
such that  when we couple a dynamical $G$ gauge field to this system,
the resulting gauged system by $Y(\mu)/G$ is
\begin{itemize}
\item  in the confined phase  in the limit $\mu \gg 0$, and
\item  in the Higgsed phase in the limit $\mu \ll 0$, 
\end{itemize}
with  no phase boundary between the two limits.
For example, when $G=\SU(N)$ we can just take $N-1$ copies of scalars in the fundamental representation. Similarly, when $G=\Sp(N)$, we can take $2N$ copies of scalars in the fundamental representation.\footnote{It would be interesting to construct such $Y(\mu)$ for other groups more explicitly. Here we consider their existence as part of the folklore theorem we rely on.}

We now consider a combined system $(X\times Y(\mu))/G$, namely, the original system $X$ together with the scalar system $Y(\mu)$ with a potential specified by a parameter $\mu$, coupled to a single $G$ gauge field.
When $\mu<0$ with $|\mu|\gg M_X$, the gauge group $G$ is completely broken in an energy scale much higher than the gap $M_X$ of the system $X$. Then we have 
\begin{equation}
\frac{X\times Y(\mu)}{G} \xrightarrow{\mu\to -\infty} X.
\end{equation}
When $\mu>0$ with $\mu\gg M_X$, the scalars in $Y$ can be integrated out in a scale much higher than the gap $M_X$ of the system, and therefore we have \begin{equation}
\frac{X\times Y(\mu)}{G} \xrightarrow{\mu\to +\infty} \frac{X}{G}.
\end{equation}

Now we see that $X$ and $X/G$ are continuously connected. 
However, for this assumption to be the case, the scalars in $Y(\mu)$ need to be able to screen all Wilson lines of the Lie algebra of $G$, because otherwise
some Wilson line shows the area law in the confining phase which can be distinguished from the Higgs phase. Thus we must impose the condition
$\pi_1(G)=0$ so that all representations of the Lie algebra of $G$ are actually allowed by the Lie group $G$.

In the rest of the section, we would like to give further credence to the discussion above, by  analyzing the case when $X$ is a system of free massive fermions more explicitly.

\subsection{Gauging free fermions}
\label{subsec:preliminary}
When $X$ is a system of free massive fermions, 
 the Lagrangian of the theory we consider on the flat space is given by 
\beq
\CL= - \ii \bar{\psi} \bar{\sigma}^\mu ( \partial_\mu +\rho(T_A)A^A_\mu)\psi- \frac{1}{2} m\left[\psi \psi +{\rm c.c.} \right]
- \frac{1}{4g^2}F^A_{\mu\nu}F^{A \mu\nu}+\frac{\theta}{64 \pi^2}\epsilon^{\mu\nu\rho\sigma}F^A_{\mu\nu}F^A_{\rho\sigma}, \label{eq:nonsusylag}
\eeq
where $\psi$ are fermions, $\rho(T_A)$ are generators of the gauge group in a representation $\rho$,
$m$ is a mass parameter, $g$ is the gauge coupling, and $\theta$ is the theta angle.
We assume that the Majorana fermions are in a strictly real representation $\rho$ of the  gauge group $G$. 
We choose the  $\sCP$ transformation to commute with the gauge symmetry.
This is possible because the representation $\rho$ is strictly real.
For more on our conventions, see Appendices \ref{sec:CPTbasics} and \ref{sec:concreteCP}.

Let us first recall the following simple fact about the chiral anomaly. 
By a change of variables $\psi=e^{\ii \alpha} \psi'$ in the path integral, 
the parameters are changed as
\beq
m'= e^{2\ii \alpha}m,~~~\theta'=\theta- 2t_\rho \alpha
\eeq 
where $t_\rho$ is an integer defined by $\tr [\rho(T_A)\rho(T_B)]=-t_\rho \delta_{AB}$,
in a normalization that
the adjoint representation has $t_\text{adj}=h^\vee$, where $h^\vee$ is the dual coxeter number of $G$.

When the mass parameter is positive and much larger than the dynamical scale of the theory, 
the IR effective theory is given by a pure Yang-Mills theory with the $\theta$ unchanged from the UV.\footnote{
It is better to regard this statement as the definition of the phase of  the fermion path integral.
In the Pauli-Villars regularization, this means that we are taking the regulator mass parameter to be positive.
}
Then, by the anomaly discussed above, we conclude that the $\theta_{\rm eff} $ in the low energy effective action in the general mass case is given by
\beq
\theta_{\rm eff}=\theta +  t_\rho \arg(m), \label{eq:efftheta}
\eeq
where $\arg(m)$ is the phase of $m$; $\arg(|m|)=0$ and $\arg(-|m|)=\pi$.

As recalled already, the system is believed to spontaneously break the $\sCP$ invariance when $\theta_\text{eff}=\pi$. 
We would like to retain the ability to change the sign of $m$ from positive to negative, keeping the fact that $\theta_\text{eff}=0$.
This requires that  $t_\rho \in 2\bZ$ .

There is another way to see the condition $t_\rho \in 2\bZ$. Let us consider a fermion mass which depends on the space coordinate $y:=x^3$, given by
\begin{equation}
\begin{cases}
m(y)>0, & (y>\epsilon) \\
m(y)<0, & (y<-\epsilon)
\end{cases}
 \label{eq:mvariation}
\end{equation}
for a small positive number $\epsilon$. 
In this situation, one manifestation of the nontrivial SPT phases is that  localized gapless Majorana fermions appear
at the boundary $y \sim 0$.

Now let us gauge the massless Majorana fermions at the boundary by a gauge group $G$ in a representation $\rho$. 
In the 3d theory, there is a parity anomaly. 
One manifestation of this anomaly is that under a gauge transformation, the fermion functional determinant 
changes the sign as $(-1)^{t_\rho n}$, where $n$ is an integer determined by the topology of the gauge transformation.
This anomaly exists when $t_\rho$ is an odd integer.  

To cancel this anomaly, we have to introduce a Chern-Simons term with half-integer Chern-Simons level.\footnote{More precisely we should use the language of the $\eta$ invariants to state what is going on~\cite{Seiberg:2016rsg}. For our purposes here, using a somewhat naive language of half-integer Chern-Simons level already implies that we need that $t_\rho$ is even, which is all we need at this point.}
We have to distinguish two cases. If the gauge field is living solely on the 3d boundary, the $\sCP$ is explicitly broken when $t_\rho$ is odd.
If the gauge field lives in the 4d bulk, the parity anomaly is cancelled by the anomaly inflow mechanism. This is because  
the theta angles on both sides of the boundary $y=0$ are different: 
$\theta_{\rm eff}=0$ on one side, and $\theta_{\rm eff}=\pi$ on the other.
However, in this case, the $\sCP$ is spontaneously broken in the region with $\theta_\text{eff}=\pi$ as discussed above, and hence we cannot apply the SPT phase classification.
Therefore, in any case, we have to impose the condition that $t_\rho $ is an even integer.

So far, we have discussed a necessary condition
\beq
t_\rho \in 2\bZ \label{eq:dynkcond}
\eeq
so that the SPT phase is not spoiled by the gauge interaction. 
In the previous section, we argued that if $G$ is further assumed to be simple, connected and simply connected, and if $\theta_\text{eff}=0$ is satisfied, the vacuum of the gauge theory is in the same SPT phase as the  the original  theory without the gauge field.
In our free fermionic case, the condition on the theta angle imposes the condition $t_\rho \in 2\bZ$.

We would like to make further checks of this conclusion  by considering the partition function of these systems  on various manifolds.
Suppose that we have a theory which has a mass gap and no topological degrees of freedom in the sense that vacuum states in the Hilbert space is one dimensional in any manifold. 
The infrared limit of such a theory is called an invertible topological field theory.
Now we consider the partition function of this  theory on a manifold $M$, where we take its metric to be extremely large.

The partition function then is given by a phase factor $Z(M) =e^{\ii \varphi}$ 
up to uninteresting contributions which can be continuously deformed to be absorbed by local gravitational counterterms. 
When $Z(M)$ are different as functions of the choice of the manifold $M$, the SPT phases are definitely different.
We stress that this criterion does not require any detail of the UV theory. 
For example, there can be strongly coupled gauge theory in the intermediate energy scale between the UV and the IR, 
as long as the IR theory is gapped and is described by an invertible field theory.
In terms of the partition function, one consequence of our claim is then as follows:
\begin{claim}
Suppose  the group  $G$ is simple, connected and simply connected and $t_\rho$ is even.
Then, in the low energy limit, the phase of the partition function of the theory of $\nu{=}\dim\rho$ free massive Majorana fermions  on a manifold $M$ is the same as that of the gauge theory \eqref{eq:nonsusylag} with $\theta=0$.
\end{claim}
In the next two subsections \ref{subsec:eta} and \ref{subsec:top}, we will establish the claim above, by first relating the phase $\arg Z(M)$ to the properties of the $\eta$ invariant, and then by studying the dependence of the $\eta$ invariant on the dynamical gauge fields.

\subsection{Partition function and the $\eta$ invariant}\label{subsec:eta}
The aim of this subsection is to reduce the computation of the phase of the partition function of the gauge theory
to a property of the eta invariant \eqref{eq:etaMA}. The property \eqref{eq:etaMA} itself will be established in the next subsection.

The partition function $Z(M)$ of the gauge theory is given as 
\beq
Z(M)= \int [{\cal D}A] Z_\psi (M,A) e^{-S_G},
\eeq
where $Z_\psi (M,A)$ is the fermion partition function and $S_G$ is the Euclidean gauge field action
\beq
S_G=\int \frac{1}{4g^2}F^A_{\mu\nu}F^{A \mu\nu} d^4x.
\eeq
Note that this gauge field action is real and positive.

Using the  4-component Majorana fermion 
\beq \Psi=\left( \begin{array}{c}
\psi_\alpha \\ \bar{\psi}^\aa
\end{array}
\right)
\eeq
 the Euclidean space Lagrangian of fermion fields can be written as
\beq
\frac{1}{2}\Psi^T C (\Slash{D}+m)\Psi,
\eeq
where $C=\diag(\epsilon^{\alpha\beta}, \epsilon_{\aa\bb})$ is the charge conjugation matrix acting on spinor indices, and $\Slash{D}=D_\mu \gamma^\mu$
is the Dirac operator. 
This form is more appropriate when we consider the Lagrangian on  unorientable manifolds.

Now we study the fermion partition function $Z_{\psi}$, which is given by
\beq
Z_\psi(M,A) = \frac{\Pf[C(\Slash{D}+m)]  }{\Pf[C(\Slash{D}+\Lambda)]}
\eeq
where $\Pf$ is the Pfaffian of the fermion functional space, and we have introduced the Pauli-Villars regulator with mass $\Lambda>0$. 
The analysis below is essentially the same as the one given in \cite{Witten:2015aba}, except that we now have a gauge field $A$.

We may define the Pfaffian in the following way. First, note that the charge conjugation matrix $C$ has the property that
\beq
 C \gamma_\mu = (C \gamma_\mu)^T= - \gamma_\mu^*C .
\eeq
Therefore, we have
\beq
(\Slash{D} +m)\Psi=(-\ii \lambda+m)\Psi \Longleftrightarrow (\Slash{D} +m)(C\Psi^*)=(- \ii \lambda+m)(C\Psi^*) \label{eq:eigendeg}
\eeq
where $\lambda$ is an eigenvalue of $\ii \Slash{D}$, and
we have used the fact that the representation $\rho(T_A)$ is real. One can check that $\Psi$ and $C\Psi^*$ transform in the same way under $\sCP$,
and hence they are sections of the same $\pin^+$ structure.
Thus if $\Psi$ is an eigenfunction, then $C\Psi^*$ is also an eigenfunction of the same eigenvalue.
Furthermore, these two eigenfunctions are guaranteed to be distinct because of the identity $C(C\Psi^*)^*=-\Psi$. 

We learned that the eigenvalues always come in pairs.
We define the Pfaffian as the product of eigenvalues, where we take one eigenvalue from each pair $(\Psi, C\Psi^*)$
of eigenfunctions. We get
\beq
Z_\psi(M,A) =  \prod_\lambda{}' \frac{ -\ii \lambda+m}{-\ii \lambda+\Lambda} 
\eeq
where the product $\prod'$ is over all the pairs $(\Psi, C\Psi^*)$ of the eigenfunctions of the Dirac operator.

Taking the argument, we have
\beq
\arg Z_\psi(M,A) =  \sum_\lambda{}' \left[  \arg\left(\frac{ \lambda +\ii m}{\lambda}\right) - \arg\left(\frac{ \lambda + \ii \Lambda}{\lambda}\right) \right] \bmod 2\pi \bZ.
\eeq
When the eigenvalue is much smaller than $|m|$, the phase of $ (\lambda + \ii m)/\lambda$ is essentially $  \frac{\pi}{2} \sign(m) \sign(\lambda)$.

The Atiyah-Patodi-Singer $\eta$ invariant is defined as follows:
\beq
\eta(M,A)= \left( \sum_{\lambda} \sign(\lambda) \right)_{\rm reg} := \frac{2}{\pi}\lim_{\Lambda \to \infty}  \sum_{\lambda} \arg\left(\frac{ \lambda + \ii \Lambda}{\lambda}\right).
\eeq
Here the sum is taken over all the eigenmodes $\Psi$, not over pairs $(\Psi, C\Psi^*)$ as we did above.
When some $\lambda$ is zero, we formally define 
$\sign(\lambda=0)=+1$ and then the $\eta$ is defined as above.

Comparing two expressions, we see that
\begin{equation}
\arg Z_\psi(M,A) = -  \frac{\pi}{4} (1-\sign(m)) \eta(M,A) +  \sum_{k \geq 1}\CO(m^{-k}) \bmod 2\pi \bZ, 
\end{equation}
where the correction terms $ \sum_{k \geq 1}\CO(m^{-k})$ go away in the limit $|m| \to \infty$.
Physically speaking, these correction terms correspond to higher dimensional operators in the low energy effective action of the gauge field 
after integrating out the massive fermion fields.
We expect that these terms can be neglected if the mass $|m|$ is much larger than the length scale of the manifold $M$ and the dynamical scale of the gauge theory. So we assume that the mass is large and we neglect the correction terms.

Now the path integral becomes
\beq
Z(M)= \int [{\cal D}A]  \exp\left[- \ii \frac{\pi}{4} (1-\sign(m)) \eta(M,A) +{\rm crr} \right]  |Z_\psi (M,A)| e^{-S_G},\label{eq:fermipath}
\eeq
where ${\rm crr}$ represents the correction terms.
We will show in the next subsection \ref{subsec:top} that if the condition \eqref{eq:dynkcond} is satisfied, the $\eta$ invariant $\eta(M,A) \bmod 4\bZ$ is independent of the gauge field $A$,
so we can write \begin{equation}
\eta(M,A)=  \nu\,\eta_0(M) \bmod 4\bZ \label{eq:etaMA}
\end{equation}
 where $\eta_0(M)$ is the $\eta$ invariant of a single Majorana fermion. 
We finally get
\beq
Z(M)=  \exp\left[ -\ii \frac{\pi}{4} (1-\sign(m)) \nu\,\eta_0(M) \right] \int [{\cal D}A]   e^{\rm crr}|Z_\psi (M,A)| e^{-S_G}.
\eeq
The factor $ |Z_\psi (M,A)| e^{-S_G}$ is manifestly positive. Therefore, when  the correction terms can be neglected, we get
\beq
\arg Z(M)=  -\frac{\pi}{4} (1-\sign(m)) \nu\,\eta_0(M) .
\eeq
This is exactly the same as in the case of free Majorana fermions.

\subsection{Topology of gauge bundles and the $\eta$ invariant}\label{subsec:top}
In this subsection, we show that the $\eta(M,A)$ is given as
\beq
\eta(M,A)=  \nu\,\eta_0(M) +2t_\rho n \bmod 4\bZ. \label{eq:etagauge}
\eeq
where $n \in \bZ$ is an integer which is essentially the instanton number of the gauge field.
The property \eqref{eq:etaMA} immediately follows, using the fact that $t_\rho\in 2\bZ$.
We show \eqref{eq:etagauge} below by combining  the Atiyah-Patodi-Singer theorem 
and and the obstruction theory.

\paragraph{The index theorem:}
The Atiyah-Patodi-Singer index theorem for a $5=4+1$ dimensional unoriented manifold
$N$ states \cite{Atiyah:1975jf,Stolz}  that the index (for $\pin^+$ structure) is given by\footnote{In general, an index can be defined if 
we have a $\bZ_2$ grading $\epsilon = \pm 1$, and a self-adjoint elliptic operator $\ii \Slash{D}$ which is odd under the $\bZ_2$ grading, i.e., $\epsilon \Slash{D}=-\Slash{D} \epsilon$.
For the 5 dimensional $\pin^+$ structure with gamma matrices $\Gamma^I~(I=1,\ldots,5)$, we use $\epsilon=\Gamma_1 \Gamma_2 \Gamma_3 \Gamma_4 \Gamma_5$
as the $\bZ_2$ grading and $\ii \Slash{D}_{\rm 5d}=\Gamma_6 \Gamma^I D_I$, where $\Gamma_6$ is an additional gamma matrix with $(\Gamma_6)^2=1$ and 
$\Gamma_6\Gamma_I+\Gamma_I \Gamma_6=0~(I=1,\ldots,5)$ so that the relation $\epsilon \Slash{D}_{\rm 5d}=-\Slash{D}_{\rm 5d} \epsilon$ is satisfied. 
A reflection $\sCR$ in a direction $\hat{n}_I$ is defined as $\Psi \to \hat{n}_I \Gamma^I \Psi$
which commutes with $\epsilon$ as it should be so that the $\bZ_2$ graded $\pin^+$ bundle is well-defined. 
In a cylinder $N=M \times \bR$ we have $\ii \Slash{D}_{\rm 5d}=\Gamma_6\Gamma_5 (\partial_5+\ii \Slash{D}_{\rm 4d})$ 
where $\ii \Slash{D}_{\rm 4d}=\ii \gamma^\mu D_\mu$ and $\gamma^\mu:=-i \Gamma_5 \Gamma^\mu~(\mu=1,2,3,4)$. 
The eta invariant for the index problem is defined by using this $\ii \Slash{D}_{\rm 4d}$ in the subspace 
$\epsilon=+1\ (\leftrightarrow \Gamma_5=\gamma_1\gamma_2\gamma_3\gamma_4)$. 
Because of the lack of perturbative anomaly in 5 dimensions, the index gets contributions only from the
boundary eta term as in \eqref{eq:APS}.
In this setup, we can also define a charge conjugation matrix $C$ such that $C \Gamma^*_I=\Gamma_I C~(I=1,\ldots,5)$ and $C \Gamma^*_6=\Gamma_6 C$
and hence $C\Psi^*$ is a section of the same bundle as $\Psi$ with the same eigenvalue.
For a more mathematical exposition, see~\cite{Stolz}. Note that $\eta_\text{there}=\eta_\text{here}/2$. 
}
\beq
{\rm Ind} \Slash{D}_{\rm 5d}(N,A_{\rm 5d}) = -\frac{1}{2}\eta(\partial N,A).\label{eq:APS}
\eeq
One can also show similarly to \eqref{eq:eigendeg} that the 5d index is an even number for Majorana fermions,
\beq
{\rm Ind} \Slash{D}_{\rm 5d}(N,A_{\rm 5d} ) \in 2\bZ .
\eeq
Using these equations, we see that  $\eta(M,A) \bmod 4\bZ$ is   a cobordism invariant. 
This can be seen by considering a 5d manifold $N$ with $\partial N = [M_1] +[ - M_2]$, where the minus sign in $[-M_2]$ is meant to reverse the $\pin^+$ structure 
of $M_2$.

\paragraph{Some obstruction theory:}
Next we need to understand the topology of gauge bundles,
which  can be understood by the obstruction theory.
Let us first recall the notion of the CW-complex for a manifold $M$.

We write the manifold $M$ as 
\beq
M=\bigcup_{i} D_i,
\eeq
where (i) $D_i \cap D_j=\varnothing$ for $i \neq j$ inside $M$, (ii) each cell $D_i$ is homeomorphic to an open disk of dimension $n(i)$, 
and (iii) the points in the closure $\overline{D}_i$ but not in $D_i$ are contained in lower dimensional cells 
\beq
 \overline{D}_i \setminus D_i \subset  \bigcup_{n(j)<n(i)} D_j.
 \eeq
For example, an $n$-dimensional sphere $S^n$
has a CW complex $S^n=D_1 \cup D_2$, where $D_1$ is a 0-dimensional point and $D_2$ is homeomorphic to an $n$-dimensional open disk, such that all points on the boundary of $D_2$ map to the single point $D_1$.

Let us define the $d$-dimensional skeleton of $M$ as
\beq
M_{d}=\bigcup_{n(i) \leq d} D_i.
\eeq
This $M_d$ is not necessarily a manifold, but it is a reasonably well-behaved topological space.

Let us take a smaller $n(i)$-dimensional disk $D'_i$ whose closure is contained inside $D_i$, i.e., $\overline{D'}_i \subset D_i$. 
Then, $M_d$ is constructed by gluing the $d$-dimensional disks $D_i~(n(i)=d)$
with the space
\beq
M'_d := M_d \setminus \bigcup_{n(i)=d} D'_i.
\eeq
The $D_i$ is homotopically equivalent to a point, while the $M'_d$ is homotopically equivalent to $M_{d-1}$,
\beq
D_i \sim \{ {\rm pt} \},~~~M'_d \sim M_{d-1},
\eeq
where $\sim$ means the homotopy equivalence.\footnote{More precisely, the situation is as follows. Let $X$ be a topological space and $Y \subset X$ its subspace.
Suppose that there exists a continuous one parameter family of maps $f_t: X \to X~( 0 \leq t \leq 1)$ such that $f_0$ is the identity map,
$f_1(X) \subset Y$ and $f_1(y) =y$ for $y \in Y$. If such $f_t$ exists, $Y$ is said to be a deformation retract of $X$. 
Now, if there is some vector bundle $E$ on $X$, we can consider a one parameter family of bundles $E_t=f^*_t E$ on $X$
such that $E_0=E$ and $E_1$ is a pull-back of a bundle $E|_Y$ on $Y$. Then the topology of the bundle $E=E_0$ is classified by the topology of $E|_Y$.
In our situation, we are using the case $(X,Y)=(D_i, \{{\rm pt}\})$ and $(M'_d, M_{d-1})$.}
The gluing region $D_i \cap M'_d$ is homotopy equivalent to a sphere $S^{d-1}$,
\beq
D_i \cap M'_d \sim S^{d-1}.
\eeq

Now we have done enough preparation to discuss the topology of $G$-bundle on $M=M_4$. 
We will use the following facts about a simple, connected, simply connected Lie group $G$:
\beq
\pi_0(G) =0,~~\pi_1(G)=0, ~~\pi_2(G)=0,~~\pi_3(G)=\bZ.
\eeq
Suppose inductively that the gauge bundle on $M_{d-1}$ can be trivialized. Then, because of the homotopy equivalence, the bundle on $M'_{d}$ is also trivial.
The disks $D_i~(n(i)=d)$ are homotopically trivial and hence the bundle on them can also be trivialized.
We construct $M_d$ by gluing $M'_{d}$ and $D_i~(n(i)=d)$. If the bundle is trivialized on each  $M'_{d}$ and $D_i~(n(i)=d)$,
the gluing of the bundle is specified by an element of $\pi_{d-1}(G)$ for each $i~(n(i)=d)$. When $d<4$, the homotopy group $\pi_{d-1}(G)$ is zero and hence
the bundle on $M_d$ is again trivial. Thus the induction continues when $d<4$.
When $d=4$, the element of $\pi_{3}(G)=\bZ$ associated to the gluing of $D_i$ can be thought of as  the instanton number localized on the disk $D_i$.
It is  clear that topologically we can gather all the instantons to a single four-dimensional disk (say $D_0$) by continuous deformation,
and define the total instanton number $n$.

If the manifold is orientable, this is the end of the classification of $G$-bundles. The $G$-bundle on $M$ is classified by the integer $n \in \bZ$ which is the instanton number.
However, if $M$ is not orientable, there is one more twist to the story. Locally on the disk $D_0$, we can define an orientation and distinguish instantons
from anti-instantons. However, globally, if we move an instanton through a path along which the orientation flips sign, an instanton comes back as an anti-instanton.
This process changes the instanton number from $n$ to $n-2$. Therefore, only the $n \bmod 2$ can be a topological invariant. 
Recall that $n$ instanton amplitude is proportional to $e^{ \ii n \theta }$. This phase factor is consistent with the mod 2 nature of $n$ only if 
$\theta$ is $0$ or $\pi$. This is precisely the same as the requirement of $\sCP$ invariance in gauge theory. In fact, 
we can put the theory on an unorientable manifold if and only if the theory has a $\sCP$ invariance.

Another way to present what we have found in this subsubsection is as follows. The obstruction theory as described here defines an analogue of the characteristic class $c_2$ of a unitary bundle for any simple, connected and simply-connected gauge bundle, which we still denote by $c_2$ by a slight abuse of the notation. This $c_2$ is a class in $H^4(M,\bZ)$. 
When $M$ is orientable, this cohomology group is $\bZ$, and then $c_2$ defines an integer-valued instanton number. 
When $M$ is unorientable, however, this cohomology group itself is $\bZ_2$, and then $c_2$ only gives us the instanton number modulo 2.

\paragraph{Derivation:}
Now we can show our crucial identity \eqref{eq:etagauge} by using the facts established above.
We have gathered instantons on a single disk $D_0$ inside the manifold $M$. Then, we can represent the manifold $M$ as a connected sum
$M \# S^4$, where $M$ and $S^4$ are connected by a tube. The $M \# S^4$ is the same as $M$ as a manifold, but we can put all the instantons
 on $S^4$. The connected sum $M \cong M \# S^4$ is equivalent to the direct sum $M+S^4$ in the cobordism group, and we can use the cobordism invariance 
 to compute the $\eta$ as
 \beq
 \eta(M,A) &= \eta(M,A_\text{trivial})+\eta(S^4, A_\text{$n$ instantons}) \bmod 4\bZ \nonumber \\
 &=\nu\,\eta_0(M)+2t_\rho n  \bmod 4\bZ,
 \eeq
where we have used the fact that $\eta \bmod 4\bZ$ in an oriented manifold is the same as the Atiyah-Singer index, which is given by $2t_\rho n$ in an $n$-instanton background in $S^4$.
Equivalently, one can also see the fact that $\eta(S^4, A_\text{$n$ instantons})=2t_\rho n \bmod 4\bZ$ from \eqref{eq:efftheta} and \eqref{eq:fermipath}.
This establishes our claim.

\subsection{Flavor symmetries and the $\eta$ invariant}\label{subsec:flavor}
Up to now, we have considered only the $\sCP$ symmetry as the protecting symmetry defining the SPT phase.
In this subsection, as an application of the analysis of the $\eta$ invariant in the previous two subsections, we consider  what happens when the theory possess other global symmetries $\sF$.
This subsection is slightly outside of the main points of this paper, and can be skipped in the first reading.

For simplicity we assume that $\sF$ commutes with $\sCP$.
Put differently, we are going to study the system as an SPT phase protected by $\sCP \times \sF$.
We assume that gauge bundle has no effect on the $\eta$ as discussed before. However, when we have a flavor symmetry $\sF$,
we can introduce a background flavor gauge field for $\sF$.
Such a background field defines a bundle which we denote as $E_\sF$. 
The $\eta(M, E_\sF)$ in general has dependence on this flavor symmetry bundle.

If $\sF$ is a Lie group that is (semi-)simple, connected and simply connected, the effect of $E_\sF$ can be classified in completely the same way as in 
the case of gauge bundle. 
We have a relation $\eta(M,E_\sF)=\nu\,\eta_0(M)+2t_\sF n  \bmod 4\bZ$ for some parameter $t_\sF$.
If $t_\sF$ is odd, then the effect of flavor bundle is nontrivial.

However, $\sF$ need not be (semi-)simple, connected or simply connected. Rather than doing a systematic analysis, let us give a simple example to illustrate the point.
Suppose that we have $\nu$ free Majorana fermions with the same mass parameter. Then the  theory has $\OO(\nu)$ flavor symmetry. 
Let us suppose that we add interactions to this system, and the symmetry is explicitly broken down to a subgroup, say $\sF=(\bZ_2)^\nu$ which acts on
each Majorana fermion as $(-1)$. 

In any unoriented manifold $M$ of spacetime dimension $d$, there is an orientation line bundle $\CE=\wedge^d T M$.
The transition function of this bundle can be taken to be $\pm 1$, and hence it is a $\bZ_2$ bundle. 
Using this bundle, we can consider a flavor bundle given by
\beq
E_\sF=\bigoplus_{i=1}^{\nu}\CE^{s_i},
\eeq
where $s_i=0$ or $1$. 

If we have a $\pin^+$ structure, then $\pin'^+:=\CE \otimes \pin^+$ is another $\pin^+$ structure which is conjugate to the original one in the sense that 
all the eigenvalues of the Dirac operator on $\pin'^+$ has the opposite sign from those of $\pin^+$.
From this fact, we can see that the $\eta$ under the above flavor bundle is given by
\beq
\eta(M,E_\sF)=\sum_{i=1}^{\nu} (-1)^{s_i}\eta_0(M).\label{eq:etaincludeflavor}
\eeq
The implication of this equation is as follows. For $\nu{=}16$, $\eta(M) \bmod 4\bZ$ is trivial if fermions are not coupled to nontrivial bundles. 
However, once we introduce a flavor bundle $E_\sF$, the $\eta(M,E_\sF)$ becomes nontrivial. This means that the boundary theory of this SPT phase
must be nontrivial, when we require that the interactions preserve the $\sF$ symmetry.

Now let us consider the case that we are not imposing $\sF$ as a symmetry protecting SPT phases, but it is just an accidental symmetry.
We denote this accidental symmetry as $F$.
Then the above discussion gives us a simple necessary criterion for an interaction that collapses the free fermionic classification:
\begin{claim}
The interaction term that collapses the free fermionic classification must be sufficiently generic so that the accidental flavor symmetry $F$ which remains unbroken by the interaction is small enough such that 
the quantity $\eta(M,E_F) \bmod 4\bZ$
does not depend on the flavor symmetry bundle $E_F$.
\end{claim}

For example, in the case of $1+1$ dimensional system of class BDI with $\nu{=}8$, 
the interactions which Fidkowski and Kitaev introduced~\cite{Fidkowski:2009dba} to gap the boundary mode breaks the symmetry from $\OO(8)$ down to $\Spin(7)$. 
The bundle of $\Spin(7)$ in two dimensions is always trivial, so it has no effect on $\eta$. Here it is important that the unbroken group is $\Spin(7)$ instead of 
$\SO(7)=\Spin(7)/\bZ_2$.

\section{QCD as SPT phases}\label{sec:lowenergy}
In the previous section, we have seen that the SPT phases of the free Majorana fermions 
do not change even if we add the gauge interaction,  as long as $t_\rho \in 2\bZ$
and the gauge group is simple, connected and simply-connected.
In this section, we study this statement from   the viewpoint of  the low energy effective theory of Goldstone bosons after the color confinement in theories of quantum chromodynamics (QCD), with gauge groups $\SU(N)$, $\Spin(N)$ and $\Sp(N)$.

\subsection{The models}
We consider $\SU(N)$ theory with $N_f$ fundamental flavors $\psi$ and $\tilde\psi$ in the representation $\rho= N_f \otimes ({\bf N}\oplus\bar{\bf N})$, $\Spin(N)$ theory with $N_f$ fundamental flavors $\psi$ in  
$\rho=N_f \otimes {\bf N}$ and $\Sp(N)$ theory with $2N_f$ half-flavors $\psi$ in $\rho=2N_f \otimes {\bf 2N}$.
Here we are considering $\Spin(N)$ rather than $\SO(N)$ to agree with our condition $\pi_1(G)=0$, but this difference
is not so important as far as the Goldstone bosons are concerned.

First, let us summarize what is believed to happen in these theories. See e.g., \cite{Weinberg:1996kr} for a standard textbook.

In the massless case, these gauge theories have flavor symmetry $F_0$ which is $F_0=\SU(N_f)_L \times \SU(N_f)_R$ in $\SU(N)$ theory,\footnote{There is also
$\U(1)$ baryon symmetry, but it is irrelevant for the discussion below and we neglect it.} 
$F_0=\SU(N_f)$ in $\Spin(N)$ theory,
and $F_0=\SU(2N_f)$ in $\Sp(N)$ theory. If we add a mass, these flavor symmetries are broken down to a subgroup.
Maximal possible flavor symmetries with massive fermions are $F=\SU(N_f)$ in $\SU(N)$ theory, $F=\SO(N_f)$ in $\SO(N)$ theory, and
$\Sp(N_f)$ in $\Sp(N)$ theory. 
It was proved, under some technical assumption by Vafa and Witten \cite{Vafa:1983tf}, that these flavor symmetries $F$ preserved by
the mass term is not spontaneously broken. 
It is also believed  that symmetries which are in $F_0$ but not in $F$ are all spontaneously broken 
when $N\gg N_f$.
Let us see each case in more detail.

The fermion Lagrangians are  given as follows: 
\begin{align}
\SU(N)&:&
\CL&= -\ii \bar{\psi} \bar{\sigma}^\mu D_\mu \psi   -i \bar{\tilde{\psi}} \bar{\sigma}^\mu D_\mu \tilde{\psi}   
-  m( \tilde{\psi}_a^i\psi^a_i + \bar{\psi}{}_a^i  \bar{\tilde{\psi}}{}^a_i ), \label{eq:suLag2}\\
\Spin(N)&:&
\CL&= -\ii \bar{\psi} \bar{\sigma}^\mu D_\mu \psi     
-  \frac{1}{2}m( \psi^a_i  \psi^a_i + \bar{\psi}{}_a^i \bar{\psi}{}_a^i   ), \label{eq:soLag2}\\
\Sp(N)&:&
\CL&= -\ii \bar{\psi} \bar{\sigma}^\mu D_\mu \psi - \frac{1}{2} m[ (J^{-1})^{ij} (J^{-1})_{ab}  \psi^a_i\psi^b_j +(J)_{ij} (J)^{ab} \bar{\psi}{}_a^i\bar{\psi}{}_b^j]. \label{eq:spLag2}
\end{align}
We took the mass matrix to be proportional to the unit matrix for $\SU(N)$ and $\Spin(N)$, and to $J$ for $\Sp(N)$.
The action of $\sCP$ is discussed in greater detail in Appendix~\ref{sec:concreteCP}, see \eqref{eq:suLag}, \eqref{eq:soLag}, \eqref{eq:spLag} in particular.

The parameter $\nu$ and the flavor symmetries in the massless case $F_0$ and the massive case $F$ are summarized in the following table:
\begin{equation}
\begin{array}{c|c|c|c}
& \nu & F_0 & F  \\
\hline
\SU(N) & 2NN_f & \SU(N_f)_L\times \SU(N_f)_R & \SU(N_f) \\
\Spin(N)& NN_f & \SU(N_f) & \SO(N_f) \\
\Sp(N)& 4NN_f & \SU(2N_f)&  \Sp(N_f)
\end{array}.
\end{equation}

When $N$ is large enough and $m$ is small enough, low energy dynamics is described by (pseudo) Goldstone bosons  associated to the spontaneous symmetry breaking from $F_0$ to $F$.
The condensate is given by 
\begin{align}
\SU(N): & \qquad\tilde{\psi}^i_a\psi^a_j=-v^3 U^i_j \label{eq:sugold}, \\
\Spin(N): & \qquad\psi^a_i\psi^a_j=-2v^3 U_{ij} \label{eq:sogold},\\
\Sp(N):& \qquad  (J^{-1})^{ik} (J^{-1})_{ab} \psi^a_k\psi^b_j=-2v^3 U^i_j \label{eq:spgold}
\end{align}
where $v$ is the mass scale of the condensate, and $U$ is the unitary matrix representing the Goldstone bosons.
The $\sCP$ transformation acts on the matrix $U$ as 
\beq
\sCP (U) =U^\dagger.
\eeq
The properties of the matrix $U$ can be summarized as follows:
\begin{itemize}
\item For $\SU(N)$, $U$ takes values in $[\SU(N_f)_L \times \SU(N_f)_R] / \SU(N_f) \simeq \SU(N_f)$ and hence it is a special unitary matrix $U^\dagger U={\bf 1}$ and $\det U=1$. 
\item For $\Spin(N)$, $U$ takes values in $\SU(N_f)/ \SO(N_f)$ and it is a special unitary matrix $U^\dagger U={\bf 1}$ and $\det U=1$ which is also symmetric $U^T=U$. 
The fact that $U$ takes values in $\SU(N_f)/ \SO(N_f)$ may be seen by writing it as $U=V V^T$, where $V \in \SU(N_f)$ 
with gauge invariance $V \sim VW$ for $W \in \SO(N_f)$.
\item Finally for $\Sp(N)$, $U'=JU$ takes values in $\SU(2N_f)/ \Sp(N_f)$ and it is a unitary matrix $U'^\dagger U'={\bf 1}$ which is anti-symmetric $U'^T=-U'$ and $\Pf(U')=1$. 
The fact that $U'$ takes values in $\SU(2N_f)/ \Sp(N_f)$ may be seen by writing it as $U'=V J V^T$, where $V \in \SU(2N_f)$ 
with gauge invariance $V \sim VW$ for $W \in \Sp(N_f)$.
\end{itemize}

When the mass is zero, these Goldstone bosons are massless, but when the mass is turned on, they have a potential energy
\beq
V_{\rm potential}=-mv^3( \tr U +\tr U^\dagger).
\eeq
If $m>0$, the vacuum is at $U={\bf 1}$. 
The number $t_\rho$ is given uniformly by $t_\rho=N_f$. From the reasons discussed in the previous section, 
we require $t_\rho\in 2\bZ$. So we restrict attention to the case $N_f$ is even.
Assuming this, when $m<0$, the vacuum is at $U=-{\bf 1}$. Notice that the condition $N_f \in 2\bZ$ is necessary from this point of view  because $\det(-{\bf 1})=(-1)^{N_f}$
for $\SU(N)$ and $\Spin(N)$, and $\Pf(-J)=(-1)^{N_f}$ for $\Sp(N)$.

\subsection{Phases from the Goldstone boson effective action}\label{sec:phasefromgoldstone}
The Wess-Zumino-Witten (WZW) term in the low energy theory of Goldstone bosons is crucial in reproducing the non-trivial value of $\arg Z(M)$ on a manifold $M$
obtained in the UV path integral argument. 
The basic properties of the WZW terms are reviewed in Appendix~\ref{sec:WZW}.
Let us now discuss concrete examples in which the low energy effective action of a strongly coupled gauge theory gives a nontrivial phase $\arg Z(M)$, reproducing the nontrivial SPT phase.

The situation is as follows. 
When the mass $m$ is nonzero, the Goldstone boson gets massive and there is a unique vacuum with unbroken $\sCP$ symmetry, since we assume  $N_f \in 2\bZ$.
 We want to compute the $\arg Z(M)$ of theories with $m>0$ and $m<0$, and see whether they  match the expectation from the UV path integral analysis.
More precisely, we consider the  difference\footnote{The value of $\arg Z(M)$ itself cannot be computed by the following reason.
From the UV point of view, the value $\arg Z(M)$ depends on the sign of the Pauli-Villars mass parameter, but that information is missing in the Goldstone boson effective action.
Also, some manifolds which give nontrivial values of $\arg Z(M)$ such as $\mathbb{RP}^4$ cannot be a boundary of any five dimensional manifold and hence there is no natural way to define the WZW term. This is one of the limitations on the low energy effective theory of Goldstone bosons. Still the difference of the phase is a perfectly well-defined quantity and can be computed using the WZW term.} of $\arg Z(M)$ for $m>0$ and $m<0$,
 \beq
 \delta \arg Z(M):= \arg Z(M)|_{m>0}-\arg Z(M)|_{m<0}
 \eeq
which can be  computed as follows. 

Let us smoothly change the mass parameter $m$ within some range $I=[-m_0, m_0]$ where $m_0>0$. 
Then  we have
\beq
 \delta \arg Z(M) =S_{\rm WZW}(I \times M)  =\int_{I \times M} I_{5}(V^{-1} dV) ,
\eeq
where the five dimensional manifold $N=I \times M$ has a boundary $\partial N=[M]+[-M]$,
and \begin{equation}
I_5=2\pi \kappa \Omega_5 = 2\pi \kappa \frac{(-1)}{(2\pi \ii)^3}\frac{2!}{5!} \tr (V^{-1}dV)^5.
\end{equation}
where $V$ takes values in the coset space $F_0/F$ and is related to $U$ as discussed above, and the value of $\kappa$ is determined by the 't~Hooft anomaly for
$(F_0)^3$.

Note that this integral over $I\times M$ makes sense even on an unorientable manifold, since $I_5$ receives an additional sign change when the orientation is reversed, due to the action of $\sCP$ on $V$.
In practice, we compute this integral by considering an oriented double cover of $M$ which we denote as $\tilde{M}$.
This $\tilde{M}$ is oriented, and $M$ is obtained from $\tilde{M}$ as $M=\tilde{M}/\bZ_2$, where the $\bZ_2$ action on the manifold is given by an orientation reversing diffeomorphism which we denote as $\sigma: \tilde{M} \to \tilde{M}$. Then we have
\beq
\delta \arg Z(M)  = \frac{1}{2}\int_{I \times \tilde{M}} I_{5}(V^{-1} dV) ,
\eeq
where the factor of $1/2$ comes from the fact that $M$ is half of $\tilde{M}$.

On the double cover $\tilde{M}$, the field $V$  must be consistent with the fact that it must  reduce to  a configuration on $M$. Denoting the coordinates of $I$ and $\tilde{M}$ as $t$ and $x$ respectively, the correct rule is that under the action of $\sigma$ it behaves as
\beq
V(t,\sigma(x))=V_\sCP (t,x),\label{eq:CPcond}
\eeq
where $V_{\sCP}$ is the $\sCP$ action on the value of $V$.
We also need to impose the condition that the values of $V$ at the boundaries of $I$ go to the vacuum expectation values 
\beq
V|_{t=m_0}= V_{{\rm vac},m>0},~~~V|_{t=-m_0}=V_{{\rm vac},m<0}.\label{eq:vaccocnd}
\eeq

Let us further restrict our attention to the case $M=\RP^4$ and $\tilde{M}=S^4$.
Because of the condition \eqref{eq:vaccocnd}, we may think of $I \times S^4$ as $S^5$ by shrinking $S^4$ at the ends of $I$.
The north pole and the south pole of $S^5$ correspond to $t=m_0$ and $t=-m_0$, respectively.
Realize $S^5$ as a unit sphere in a flat six-dimensional space with coordinates $X^I$ ($I=1,\ldots,6$), and let $\hat{X}^I$ be the points on $S^5$ with $(\hat{X})^2=1$.  Let $\hat{X}_0$ be the north pole. The action of $\sigma$ is then given by
\beq
\sigma(\hat{X})= -\hat{X}+2\hat{X}_0 (\hat{X}_0 \cdot \hat{X}) .
\eeq
This action fixes the north pole $\hat{X}=\hat{X}_0$ and the south pole $\hat{X}=-\hat{X}_0$.
We  now compute $\delta \arg Z(M)$ in some specific examples.

\paragraph{$\SU(N)$ with $N_f{=}4$.} In this case, we can take $V=(V_1,V_2) \in \SU(N_f) \times \SU(N_f)_R$ and $U=V_1V_2^{-1}$. 
In terms of these variables, the WZW term is given by
\beq
S_{\rm WZW}=\pi  N \cdot \int_{S^5} \Omega_{5}(U),
\eeq
where we used the fact that $\kappa=N$.

We consider the configuration \eqref{eq:winding1} 
\beq
U=P_{+}(\Gamma \cdot \hat{X}_0)(\Gamma \cdot \hat{X}) \label{eq:Uconfig}
\eeq
where $\Gamma^{I=1,\ldots,6}$ are the $8\times 8$ Gamma matrices and $P_+$ is the projection to the positive chirality space.
Because of this projection, the right-hand-side of \eqref{eq:Uconfig} can be regarded as a $4 \times 4$ matrix, suitable for  $N_f=4$. 
This configuration has exactly the desired properties: it has the vacuum values $U(\hat{X}_0)={\bf 1}$ and $U(-\hat{X}_0)=-{\bf 1}$ at the north pole and the south pole,
and it satisfies the condition \eqref{eq:CPcond} where $U_\sCP=U^\dagger$. 
The computation of the WZW action is reviewed in \eqref{eq:topinv}, with the result
\beq
\delta \arg Z(\RP^4) =\pi N= \frac{2\pi}{16} \nu ,
\eeq
where we have used $\nu{=}2N_fN=8N$. By using the fact that $\eta(\RP^4)=-1/4$, this result exactly reproduces the path integral computation.

\paragraph{$\Spin(N)$ with $N_f=8$.} In the $\Spin(N)$ gauge theory, we have $U=VV^T$. In terms of this variable, the WZW term is given as
\beq
S_{\rm WZW}= \frac{1}{2} \pi  N \cdot \int_{S^5} \Omega_{5}(U),
\eeq
where we have used the fact that $\kappa=N$, $\int \Omega_{5}(VV') = \int \Omega_{5}(V)+ \int \Omega_{5}(V')$ for a closed manifold, 
$\int  \Omega_{5}(V^T)=\int \Omega_{5}(V)$ and hence $\int \Omega_5(U)=2\int \Omega_5(V)$. 

We want to consider a configuration like \eqref{eq:Uconfig}, but we cannot directly use it because $U$ must satisfy the condition $U^T=U$.
Instead, we can consider a configuration
\beq
U &=\frac{1}{2}\left( \begin{array}{cc}
1 & -\ii  \\
-\ii  & 1
\end{array} \right)
\left( \begin{array}{cc}
P_{+}(\Gamma \cdot \hat{X}_0)(\Gamma \cdot \hat{X}) & 0 \\
0 & (P_{+}(\Gamma \cdot \hat{X}_0)(\Gamma \cdot \hat{X}))^T
\end{array} \right) \left( \begin{array}{cc}
1 & \ii  \\
\ii  & 1
\end{array} \right) \nonumber \\
&=\left( \begin{array}{cc}
S & \ii A \\
-\ii A & S
\end{array} \right)
\eeq
where $S$ and $A$ are the symmetric and antisymmetric part of $P_{+}(\Gamma \cdot \hat{X}_0)(\Gamma \cdot \hat{X}) $, respectively.
This is possible for $N_f=8$. One can check that it satisfies the desired properties.
Then we get 
\beq
\delta \arg Z(\RP^4) =\pi N= \frac{2\pi}{16} \nu ,
\eeq
where $\nu{=}N_f N=8N$. This again reproduces the phase we determined from the UV the path integral.

\paragraph{$\Sp(N)$ with $N_f=2$.}
In the $\Sp(N)$ theory, we define $U=J^{-1}U'=J^{-1} V J V^T$ and get
\beq
S_{\rm WZW}=\pi  N \cdot \int_{S^5} \Omega_{5}(U),
\eeq
where we have used $\kappa=2N$ and $\int \Omega_5(U)=2\int \Omega_5(V)$.

The $U$ must satisfy $(JU)^T=-JU$. Actually, it turns out that we can just use \eqref{eq:Uconfig} in this case.
The $6$ dimensional Clifford algebra $\Gamma^I$ has a charge conjugation matrix $C$ which has the property that $(C\Gamma_M)^T=-C\Gamma_M$ and 
$[C\Gamma_M, P_+]=0$.
Then, we can identify $J$ as $J=C\Gamma \cdot \hat{X}_0$. This gives the desired property $(JU)^T=-JU$.
Therefore we get
\beq
\delta \arg Z(\RP^4) =\pi N= \frac{2\pi}{16} \nu ,
\eeq
where $\nu{=}4N_f N=8N$. Again, we see that the UV phase is reproduced from the WZW term.

\paragraph{Other values of $N_f$?} The reader may have noticed that in all the examples above, $\nu$ is a multiple of $8$.
There is a reason for this. In the case of $M=\RP^4$ and $\tilde{M}=S^4$, we have argued that $\delta \arg Z(M)$ is given as
\beq
\delta \arg Z(M)=\frac{1}{2} S_{\rm WZW}(S^5),
\eeq
where the factor $1/2$ came from replacing $M$ to $\tilde{M}$.
Now, the consistency of WZW term requires that $S_{\rm WZW}(S^5)$ is in any case an integer multiple of $2\pi$.
Therefore, we can only get an integer multiple of $\pi$ for $\delta \arg Z(M)$ in the computation involving Goldstone bosons alone.
This is consistent with the known fact \cite{Wang:2014lca,Metlitski:2014xqa} that the free fermionic phases when $\nu$ is a multiple of $8$ is in fact a bosonic SPT, and that the our low-energy theory is described just by the Goldstone modes.
More precisely, we can match the anomaly by the WZW term if $N_f$ is a multiple of $4$ for $\SU(N)$, a multiple of $8$ for $\Spin(N)$, and a multiple of 2 for $\Sp(N)$.

What happens in other cases, such as the $\SU(N)$ theory with $N_f=2$? In those cases, we cannot find a configuration of $U$ which satisfies the conditions
\eqref{eq:CPcond} and \eqref{eq:vaccocnd}. Then, the computation of $\delta \arg Z(M)$ is not possible within the low energy effective theory of Goldstone bosons.
This means that somewhere on the manifold $M$ the system is forced to be out of the IR limit,
and we will have to take into account other massive excited states (hadrons) to compute $\delta \arg Z(M)$. 
It just represents a limitation of the low energy effective theory for light degrees of freedom and should not be regarded as any illness of the theory.
For example, if we consider $\SU(2)$ theory with one fermion in the adjoint representation (which is the softly broken $\CN=1$ Super-Yang-Mills), 
the $\nu=3$ SPT phase is realized. In that case,
there is no Goldstone boson at all and 
obviously we need more than low energy effective theory of light degrees of freedom.

It might be interesting to tackle the computation involving massive hadrons 
in a calculable framework such as supersymmetric domain walls~\cite{GaiottoTalk} or  holographic approaches to QCD. 
But these are beyond the scope of this paper.

\paragraph{Massless excitations on the boundary.}
Before moving on, let us see what we get at the boundary in the low energy effective theory
between the regions $m>0$ and $m<0$. We will see that the boundary theory is purely bosonic without any fermions.
Recall that in the region $y>\epsilon$ where $m(y)>0$, the vacuum is at $U=\mathbf{1}$, and that in $y<-\epsilon$, the vacuum is $U=-\mathbf{1}$.

A minimal energy path connecting these two regions would be given by
\beq
U(y) =\diag ( e^{\ii \rho(y)}, \cdots, e^{\ii \rho(y)}, e^{-\ii \rho(y)}, \cdots, e^{-\ii \rho(y)}),
\eeq
where $\rho(y)$ is a smooth monotonic function with $\rho(y \to \infty)=0$ and $\rho(y \to -\infty)=\pi$, and there are equal numbers of eigenvalues $e^{\ii \rho(y)}$
and $e^{-\ii \rho(y)}$. This is valid for all $\SU(N)$, $\Spin(N)$ and $\Sp(N)$ theories.
The flavor symmetry is broken as follows: 
\begin{align}
\SU(N)&: & \SU(N_f) &\to \mathrm{S}[\U(N_f/2) \times U(N_f/2)],\\
\Spin(N)&: & \SO(N_f) &\to \mathrm{S}[\OO(N_f/2) \times \OO(N_f/2)], \\
\Sp(N)&: & \Sp(N_f) &\to \Sp(N_f/2) \times \Sp(N_f/2).
\end{align}
 Recall that we have to impose $N_f \in 2\bZ$, so the appearance of $N_f/2$ makes sense.
 
Therefore, the boundary theory is a non-linear sigma model whose target spaces are Grassmannians given as follows:
\begin{align}
\SU(N)&: & {\rm Gr}(N_f/2, \bC^{N_f})&=\frac{\SU(N_f)}{\mathrm{S}[\U(N_f/2) \times U(N_f/2)]},\\
\SO(N)&: & \tilde{\rm Gr}(N_f/2, \bR^{N_f})&=\frac{\SO(N_f)}{\mathrm{S}[\OO(N_f/2) \times \OO(N_f/2)]},\\
\Sp(N)&: & {\rm Gr}(N_f/2, \bH^{N_f})&=\frac{\Sp(N_f)}{\Sp(N_f/2) \times \Sp(N_f/2)} .
\end{align}
Note that  the $\sCP$ symmetry is spontaneously broken on any given point of these Grassmannians. 
This is due to the fact that it was impossible to connect $U={\bf 1}$ and $U=-{\bf 1}$
without breaking $\sCP$ in these  models. A $\sCP$ invariant configuration must satisfy $U=U^\dagger$ or equivalently $U^2=1$ which requires that all
the eigenvalues are $\pm 1$. Therefore the vacua with $U={\bf 1}$ and $U=-{\bf 1}$ cannot be connected while preserving $\sCP$.

\section{SUSY, S-duality, and the collapse of free SPT classification}\label{sec:super}

In the non-supersymmetric models described in the previous section, we have reproduced, at least when $\nu$ is a multiple of 8, the
non-trivial phase $\arg Z(\RP^4)=\pi$ when $\nu{=}8 \bmod 16$, and got a trivial phase when $\nu{=}0 \bmod 16$. 
However, we did not directly  show that the models with $\nu{=}16$ give a trivial SPT phase,
since  during the transition of the vacuum from $U=+{\bf 1}~(m>0)$ to $U=-{\bf 1}~(m<0)$ 
the $\sCP$ was spontaneously broken.

In this section, we would like to demonstrate that the case $\nu{=}16$ gives the trivial SPT phase after the introduction of interactions, by showing that we can continuously change the mass $m$
from positive to negative values while maintaining mass gap and without breaking $\sCP$. 
Our main trick is to embed the $\nu{=}16$ fermions
into \Nequals2  supersymmetric $\SU(2)$ gauge theory with $N_f{=}4$ flavors.
From the analysis presented below, it will be clear that this gapped boundary is topologically trivial.

In Sec.~\ref{subsec:model}, we first recall the basics of the supersymmetric model and its S-duality. 
Then in Sec.~\ref{subsec:dualCP}, we discuss how the $\sCP$ action on the dual side can be identified.
In Sec.~\ref{subsec:vacua}, we recall the structure of the vacua of the \Nequals2 supersymmetric model and of the model broken to \Nequals1 by an explicit mass term.
After these preparations, we construct a continuous pass from the $\nu{=}16$ phase to the $\nu{=}0$ phase in Sec.~\ref{subsec:continuous}.
In Sec.~\ref{subsec:nu8}, we discuss an essential difference we encounter when we try to perform a similar analysis for the $\nu{=}8$ case, or equivalently the $N_f=2$ model.
Finally in Sec.~\ref{subsec:pin}, we study the modification of the $\pin$ structure on the dual side. 

In this section, we mostly use physicists' notation for groups where we do not distinguish two groups with the same Lie algebras unless otherwise stated.
In the last subsection~\ref{subsec:pin}, these distinctions  become very important, and we pay due attention to them.

\subsection{The model and the S-duality}\label{subsec:model}
We consider \Nequals2  supersymmetric $\SU(2)$ gauge theory with four flavors $N_f{=}4$.
This theory itself does not have a  mass gap, so we have to break supersymmetry to obtain a unique vacuum with mass gap.
For now, let us review the situation with \Nequals2 supersymmetry.
See e.g. \cite{WessBagger,Weinberg:2000cr} for basic features of $\CN=2$ supersymmetry such as the explicit Lagrangian and $\SU(2)_R \times \U(1)_R$ R-symmetries,
and e.g. \cite{Tachikawa:2013kta} for more advanced properties.

\paragraph{The model:}
In the \Nequals1 supersymmetric language, 
the system has an $\SU(2)$ vector multiplet $V$,
a chiral multiplet $\Phi$ in the adjoint of $\SU(2)$,
and four pairs of quark superfields $Q_i,\tilde Q^i$ in the doublet of $\SU(2)$.
The \Nequals2 vector multiplet consists of $V$ and $\Phi$,
and the superfields $Q_i$ and $\tilde Q^i$ form \Nequals2 hypermultiplets.

In components, $V$ consists of a gauge field $A$ and an adjoint gaugino $\lambda_1$,
and $\Phi$ consists of an adjoint scalar also denoted as $\Phi$ and a second adjoint gaugino $\lambda_2$.
The quark superfields $Q_i$, $\tilde Q^i$ contain the scalar components again denoted by $Q_i$, $\tilde Q^i$ and the Weyl fermions $\psi_i$ and $\tilde \psi^i$.
Note that in total, there are 16 Majorana fermions in $\psi_i$ and $\tilde \psi^i$. This is the starting point of our analysis.

We use the convention that the kinetic term of $\Phi$ has a factor $1/g^2$ in front, where $g$ is the gauge coupling.
We set the theta angle to be zero, so that the system can have $\sCP$ symmetry.
The superpotential  is given by
\beq
W= \sum_{i=1}^{4} (-\tilde{Q}^i \Phi Q_i + m_i \tilde{Q}^i  Q_i )
\eeq
where we have suppressed the gauge indices. 

\paragraph{The S-duality:}
This theory is known to have a strong-weak duality, which is usually simply  called the S-duality \cite{Seiberg:1994aj}. 
Under this S-duality, the original electric theory is mapped to a dual magnetic description,
which is still given by the \Nequals2 $\dualSU(2)$ gauge theory with $N_f{=}4$ flavors.
The coupling constants of the original theory $g$ and the dual theory $\dualg$ are related as $g \sim 1/\dualg$, where we have taken the theta angle to be zero.

In the following, we distinguish the objects in the original electric side and those in the dual magnetic side by using serif fonts for the former and using sans-serif fonts for the latter.
For example, the original quarks are $Q$, the adjoint field is $\Phi$, the electric coupling in $g$, whereas the dual quarks are  $\dualQ$, the dual adjoint field is $\dualPhi$, and the dual coupling is denoted by $\dualg$.

\paragraph{Vector multiplets:}
To state the S-duality in slightly more detail, consider the situation where 
the original $\SU(2)$ is Higgsed to $\U(1)$ by a vev of the adjoint field $\Phi$.
In the dual description, the dual $\dualSU(2)$ is also Higgsed by the dual adjoint field  $\dualPhi$ to $\dualU(1)$. 
Then, the field strengths of the original $\U(1)$ and the dual $\dualU(1)$ are related by the standard electromagnetic duality. 
Namely, if we denote the electric and the magnetic fields of the original $\U(1)$ as $\vec E$ and $\vec B$, and similarly those of the dual $\dualU(1)$ by $\vec{\dualE}$ and $\vec{\dualB}$, we have the relation \begin{equation}
\vec{\dualE}=\vec B,\qquad \vec{\dualB}=-\vec E.\label{eq:EB}
\end{equation}
Note that the dual quarks $\dualQ_i$, $\tilde{\dualQ}^i$ are electrically charged under the dual gauge group. This means that they correspond to magnetic monopoles in the original electric description.

Let us say $\Phi=\diag(a,-a)$. Then,
the masses of the supersymmetric particles are known to be given by the absolute value of a rational linear combination of 
$a$, $\duala \simeq (4 \pi \ii /g^2) a$, and $m_i$.
The vev of $\dualPhi$ is then given by $\diag(\duala, -\duala) \propto \diag(\ii a,-\ii a)$, which means that we have \begin{equation}
u:=\frac{1}{2}\Tr \Phi^2  \propto -\Tr \dualPhi^2
\end{equation} up to a positive real proportionality coefficient (and an additive shift which arises at the quantum level depending on $m_i$).

This factor of $\ii$ between the vev of $\Phi$ and the vev of $\dualPhi$ can also be understood as follows.
From \eqref{eq:EB}, we know that
when the original $\vec E$ is a polar vector the dual $\vec{\dualE}$  is an axial vector.
This means that the $\sCP$ action on the dual $\dualU(1)$ has an additional minus sign.
Let us represent $\Phi=\ii \tau_A \Phi^A$, where $\tau_A~(A=1,2,3)$ are Pauli matrices and the factor of $\ii$ here was introduced to agree with our convention
in this paper that gauge generators are anti-hermitian. 
The scalar in the unbroken $\CN=2$ $\U(1)$ vector multiplet is then $\Phi^{\U(1)} :=\ii \Phi^{A=3}$.
Now, suppose the $\sCP$ acts on it as $\sCP(\Phi^{\U(1)})=+\bar\Phi{}^{\U(1)}$.
Then it acts on the dual $\dualPhi^{\dualU(1)}$ as $\sCP(\dualPhi^{\dualU(1)})=-\bar\dualPhi{}^{\dualU(1)}$ because of the additional minus sign.
This implies that, when the $\sCP$-compatible vev of $\Phi$ is $\diag(a,-a)$ with a real $a$,
the $\sCP$-compatible vev of $\dualPhi$ is $\diag(\ii a,-\ii a)$, again with a real $a$.
On the other hand, if $a$ is pure imaginary, the unbroken $\sCP$ should be defined as
 $\sCP(\Phi^{\U(1)})=-\bar\Phi{}^{\U(1)}$ and $\sCP(\dualPhi^{\dualU(1)})=+\bar\dualPhi{}^{\dualU(1)}$ which differs from the above $\sCP$
 by Weyl reflection of the $\SU(2)$ or $\dualSU(2)$ gauge group.

\paragraph{Hypermultiplets:}
When the original quarks $Q_i, \tilde{Q}^i$ have masses $m_1, m_2, m_3, m_4$, it is known that the dual quarks $\dualQ_i, \tilde{\dualQ}^i$ have masses
\beq
\dualm_1&=\frac{m_1+m_2+m_3+m_4}{2}, &
\dualm_2&=\frac{m_1+m_2-m_3-m_4}{2},\nonumber \\
\dualm_3&=\frac{m_1-m_2+m_3-m_4}{2}, &
\dualm_4&=\frac{m_1-m_2-m_3+m_4}{2} . \label{eq:Sdualmass}
\eeq

In the following we consider a simple choice of the mass terms on the electric side given by  $m_i=m~(i=1,2,3,4)$.
Then, on the dual magnetic side, we have
\beq
\dualm_1=2m ,~~\dualm_2=\dualm_3=\dualm_4=0.
\eeq
The superpotential of the dual side is then given by 
\beq
W= \sum_{i=1}^{4} (-\tilde{\dualQ}^i \dualPhi \dualQ_i ) +   2m \tilde{\dualQ}^1  \dualQ_1.
\eeq

\subsection{The $\sCP$ transformation}\label{subsec:dualCP}
To study our system as a topological phase of matter,
we need to understand the actions of the $\sCP$ transformation.
Let us study it on both sides of the duality.

\paragraph{On the original electric side:}
The $\sCP$ transformation in the original electric description acts in a simple manner:
\beq
\sCP (Q_i )= \bar{\tilde Q}_i,~~~\sCP (\tilde{Q}^i ) = \bar{ Q}^i,~~~\sCP (\Phi ) =  {\Phi}^\dagger,
\eeq
where the dagger $\dagger$ on $\Phi$ signifies the adjoint of the matrix $\Phi=\ii \tau_A \Phi^A$, meaning that we take complex conjugate and transpose.
For more details, see Appendix~\ref{subsec:susyLag}. 

Note that $Q_i$ and $\tilde Q^i$ ($i=1,\ldots,4$) belong to the same representation of the gauge group, and can be combined to a single object $\underQ_I$, ($I=1,\ldots, 8$), by defining
\beq
\underQ^a_{2i-1} = \frac{Q^a_i -  \ii J^{ab}\tilde{Q}_b^i}{\sqrt{2}},~~~\underQ^a_{2i} = \frac{\ii Q^a_i -  J^{ab}\tilde{Q}_b^i}{\sqrt{2} }.
\eeq
where $J^{ab}$ is the $\SU(2) \simeq \Sp(1)$ invariant tensor.
In this description, the lowest scalar components of $\underQ^a_I$ and $J^{ab}\bar{\underQ}_b^I$ form the $\SU(2)_R$ doublet.
The $\sCP$ transformation in terms of $\underQ$ is then 
\beq
\sCP (\underQ^a_I )=J^{ab} J_{IJ} \bar{\underQ}_b^J
\eeq
where $J_{2i-1, 2i}=-J_{2i,2i-1}=1$.

Note that the massless theory preserves $\SO(2N_f)$ flavor symmetry acting on the index $I$, but that  only the $\U(N_f) \subset \SO(2N_f)$ flavor symmetry 
commutes with the above definition of $\sCP$. 
This is because the $\sCP$ transformation involves $J_{IJ}$, which determines a complex structure on  the flavor indices.

\paragraph{On the dual magnetic side:}
We need to identify  the $\sCP$ action on the S-dual side of the theory.
This can be done by first finding a transformation $\sPinv $ which transforms simply under the S-duality.
For this purpose, it is convenient to notice that the Lagrangian of \Nequals2  theory can be obtained by dimensional reduction of 5d \Nequals1 
Lagrangian. In 5d, we have a Lorentz transformation which is an element of $\SO(1,4)$ given as $\diag(1,-1,-1,-1,-1)$. This reduces 
to a parity transformation which we denote as $\sPinv $. This definition of $\sPinv $ makes it clear that it commutes with gauge symmetry
$G$, the $\SU(2)_R$ symmetry and the flavor symmetry $\SO(2N_f)$, all of which are manifest in the 5d. 
Note that $\U(1)_R$ is not manifest in 5d, so $\sPinv $ does not have to commute with $\U(1)_R$.
In fact $\sPinv$ inverts elements of $\U(1)_R$.

Now,  a parity transformation which commutes with $G \times \SU(2)_R \times \SO(2N_f)$ is unique up to phase rotation. Indeed, suppose that there is another $\sPinv '$
with the same property. Then $\sPinv^{-1} \sPinv '$ is an internal symmetry which commutes with all the gauge and global symmetries.
The only such  symmetry in our theory (in the massless limit) is the $\U(1)_R$.
This ambiguity due to this $\U(1)_R$ phase rotation can be rotated away.
Indeed,  by using the fact that the generator of
$\U(1)_R$ anti-commutes with $\sPinv $ in 4d, the phase can be essentially eliminated by redefining the phases of fields by $\U(1)_R$.
Therefore, $\sPinv $ is essentially unique, and so it is invariant under the S-duality\footnote{More precisely, we also need to worry that the center of $G\times \SU(2)_R\times \SO(2N_f)$ might mix in $\sPinv$.   Here we show that the effect of centers of $\SU(2)_R$ and $\SO(2N_f)$ can be ascribed to that of the center of $G=\SU(2)$. Indeed, the element $-1\in \SO(2N_f)$ only acts as $-1$ on the quarks, on which this is equal to $-1\in G$. Also, the element $-1\in \SU(2)_R$ acts as the product of $-1\in G$ and 
the fermion number $(-1)^F$. But $(-1)^F$ is actually an element of $\U(1)_R$ which is taken care of in the main text.   
The center of $G$ will be dealt with later in \eqref{eq:uptocenter} and in \eqref{eq:modifiedCP}; at that point, we need to determine how much gauge transformation to mix to $\sPinv$ in any case. }. 

From the 5d construction, we can see that $\sPinv $ acts on scalars as
\beq
\sPinv  (\underQ^a_I) = \underQ^a_I,~~~\sPinv  (\Phi^A) = - \bar{\Phi}^A.
\eeq
Here, we slightly abused the notation to refer to the scalar component of a superfield by the same symbol.
We chose the $\U(1)_R$ rotation so that ${\rm Re}(\Phi^A)$ and ${\rm Im}(\Phi^A)$ come from the  vector  field and the scalar field in 5d, respectively.

Two parity transformations $\sPinv$ and $\sCP$ differ by an ordinary internal symmetry.
We find that we have \begin{equation}
\sCP= \sC_{\SU(2)R} \sC_{\SO(8)} \sPinv
\end{equation}
where $\sC_{\SU(2)R} \in \SU(2)_R$ is such that
\beq
\sC_{\SU(2)R} (\underQ^a_I ) = J^{ab} \bar{\underQ}^I_b
\eeq
and $\sC_{\SO(8)} \in \SO(8)$ is such that
\beq
\sC_{\SO(8)} (\underQ^a_I)  =  J_{IJ} \underQ^a_J.
\eeq
Here notice that $I, J$ are indices of $\SO(2N_f)$, and hence it does not matter whether they appear as superscript or subscript.
In terms of  the variables $Q^a_i$ and $\tilde{Q}_a^i$, the $\sC_{\SU(2)R} $ and $\sC_{\SO(8)}$ acts as
\beq
\sC_{\SU(2)R} (Q^a_i ) = -\ii \bar{\tilde Q}^a_i ,~~~\sC_{\SU(2)R} (\tilde{Q}^a_i ) = \ii \bar{Q}^a_i \\
\sC_{\SO(8)} (Q^a_i ) =  \ii Q^a_i,~~~~\sC_{\SO(8)} (\tilde{Q}_a^i )=-\ii \tilde{Q}_a^i.
\eeq

As discussed above, the action of $\sPinv $ in the dual can be readily identified.
The action of $\sC_{\SU(2)R}$ in the dual is also easy to find, since it is just an element of $\SU(2)_R$.

It takes some work to determine the action of $\sC_{\SO(8)}$ on the dual side.
The flavor symmetry $\SO(8)$ has a subgroup
\beq
\SU(2)_1 \times \SU(2)_2 \times \SU(2)_3 \times \SU(2)_4   \subset \SO(4) \times \SO(4) \subset \SO(8) . \label{eq:foursu2}
\eeq
Let $h_p~(p=1,2,3,4)$ be the Cartan generators of $\SU(2)_p$ normalized so that fundamental representations of $\SU(2)_p$ have $h_p=\pm 1/2$. 
Then we have
$
\sC_{\SO(8)}= \exp[\pi \ii (h_1+h_3)].
$
Under the S-duality, it is known that $\SU(2)_2$ and $\SU(2)_3$ are exchanged \cite{Seiberg:1994aj,Gaiotto:2009we}, see also Appendix~\ref{sec:Sd}) for more details. 
Let $\dualh_p$ be the corresponding generators on the dual side.
Then we get
\beq
\sC_{\SO(8)}= \exp[\pi \ii (h_1+h_3)]=\exp[\pi \ii (\dualh_1+\dualh_2)]=\sC_{\dualSO(8)}
\eeq
Denoting the dual quarks as $\underdualQ_I=(\dualQ_i,\tilde{\dualQ}^i)$, we now see
\begin{align}
\sC_{\dualSO(8)}(\underdualQ_I)&=-\underdualQ_I, & (I&=1,2) \nonumber \\
\sC_{\dualSO(8)}(\underdualQ_I)&=+\underdualQ_I. & (I&=3,4,5,6) \label{eq:dualCSO8}
\end{align}

Let us check the transformation of mass parameters \eqref{eq:Sdualmass}. In $\CN=2$ supersymmetry,
the operators $(J^{-1})_{ab} S^a_I S^b_J$, which is in the adjoint representation of $\SO(8)$, are in the same supermultiplet as the conserved current of $\SO(8)$.
In particular, there is a correspondence between the symmetry generators $h_i$ and operators $\tilde{Q}^i Q_i$ as
\beq
h_1: \tilde{Q}^1 Q_1+\tilde{Q}^2 Q_2,~~~h_2:\tilde{Q}^1 Q_1-\tilde{Q}^2 Q_2, \nonumber \\
h_3: \tilde{Q}^3 Q_3+\tilde{Q}^4 Q_4,~~~h_2:\tilde{Q}^3 Q_3-\tilde{Q}^4 Q_4,
\eeq
Notice that $h_1+h_3$ corresponds to $\sum_{i=1}^4 \tilde{Q}^i Q_i$ which is associated to the overall $\U(1)$ rotation of quarks.
By the S-duality which exchanges $h_3$ and $\dualh_2$, the operators $\tilde{Q}^3 Q_3+\tilde{Q}^4 Q_4$ and
$\tilde{\dualQ}^1 \dualQ_1-\tilde{\dualQ}^2 \dualQ_2$ are exchanged. Therefore, the mass term $m \sum_{i=1}^4 \tilde{Q}^i Q_i$ goes to 
$2m \tilde{\dualQ}^1 \dualQ_1 $. More general mass parameters can be treated in the same way.

The analysis so far fixes the action of $\sCP$ on the gauge invariant operators on the dual side:
\begin{equation}
\sCP\sim \sC_{\SU(2)R} \sC_{\SO(8)} \sPinv\label{eq:uptocenter}
\end{equation}
where  we used the symbol $\sim$ to emphasize that we  still need to  determine how to  mix the $\dualSU(2)$ gauge transformation.
The precise dual $\dualSU(2)$ transformation needed can be fixed by demanding that they preserve the vevs of the various fields which we will determine later, in  \eqref{eq:adjvev}, \eqref{eq:qvev1}, \eqref{eq:qvev2}.
We also come back to this issue in Sec.~\ref{subsec:pin} where we treat the formulation of the dual theory on an unoriented manifold more carefully.

\subsection{The structure of the vacua}\label{subsec:vacua}
\paragraph{With \Nequals2 supersymmetry:}
We first discuss the \Nequals2  supersymmetric case where $m_\Phi=0$, or equivalently, $m_\dualPhi=0$. There is a moduli space of vacua spanned by $u:= \Tr \Phi^2\propto -\Tr \dualPhi^2$.
This moduli space is called the $u$-plane. On this $u$-plane, there are three distinguished points which we call A, B and C.
The interpretation of these points depends on whether we are in the original electric theory  or in the dual magnetic theory.

\begin{figure}
\centering
\begin{tabular}{c@{\qquad}c}
\includegraphics[width=.45\textwidth]{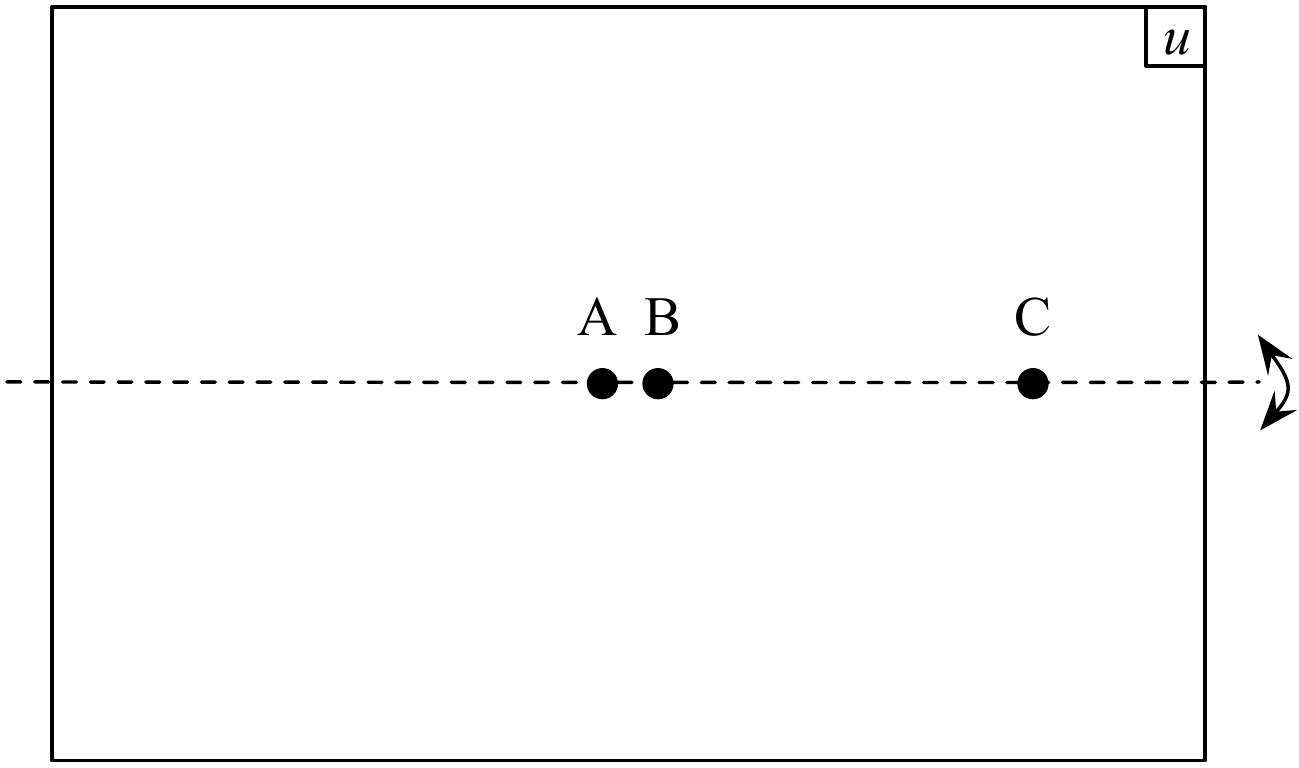}&\includegraphics[width=.45\textwidth]{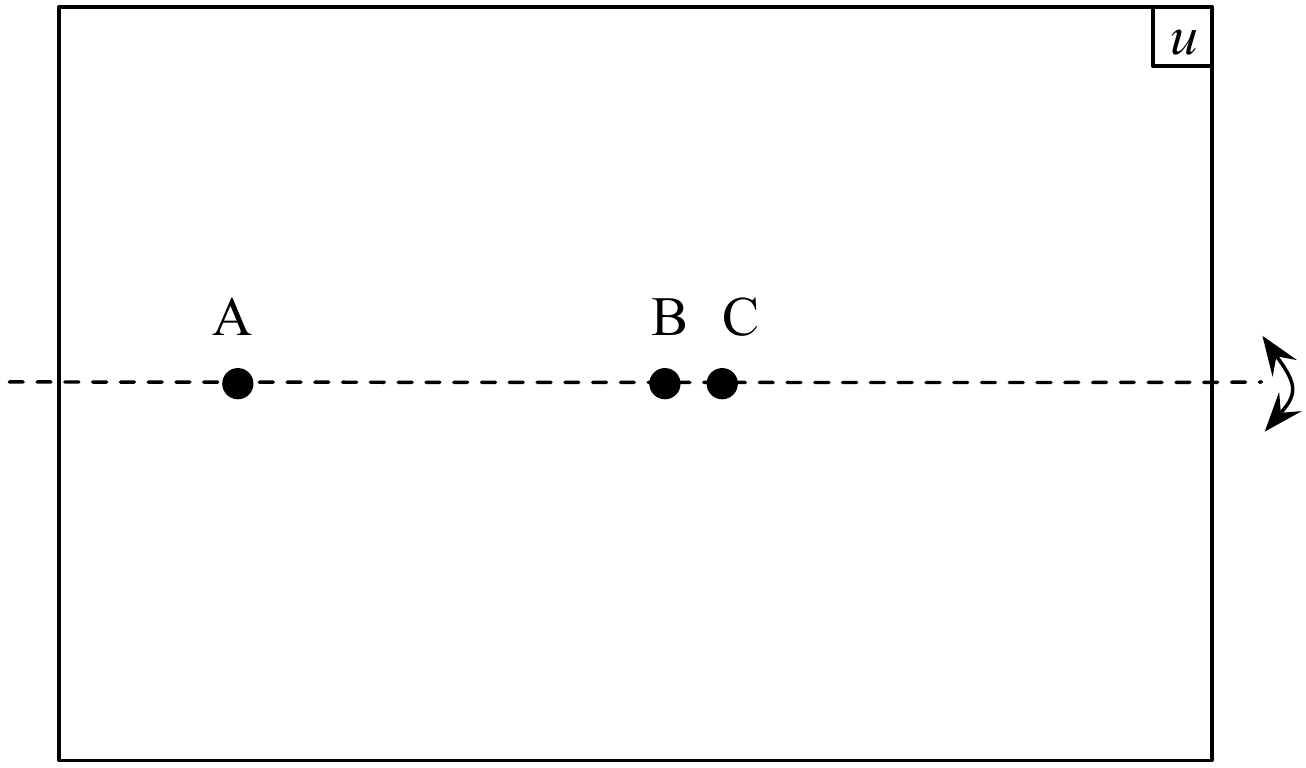}\\
Electric side & Magnetic side
\end{tabular}
\caption{On the electric side, the point $C$ is where the four quarks become massless, while the point $A$ and $B$ are the points where a monopole and a dyon become massless, respectively.
On the magnetic side, the point $A$ is where a dual quark becomes massless, the point $B$ is where the dyon becomes massless, and the point $C$ is where four monopoles become massless. The $\sCP$ acts by sending $u\to \bar u$. In particular, the points $A$, $B$ and $C$ preserve $\sCP$ and are on the real axis. \label{fig}}
\end{figure}

First consider the original electric theory.
For an illustration, see the left hand side of Figure~\ref{fig}.
We have a perturbative vacuum $\Phi=\diag(m, -m)$ where the quarks become massless. 
We denote this vacuum as C.
The other special vacua are realized in a region where the vev of $\Phi$ is small
and the system is strongly coupled.
In such vacua, we can integrate out massive quarks $Q, \tilde{Q}$ and get \Nequals2  pure Super-Yang-Mills (SYM).
In this pure SYM, there is a point where a magnetic monopole becomes massless and another point where a dyon becomes massless.
We call the former point  A and the latter point B.

In the dual magnetic theory, the points A, B and C have different interpretation. 
See the right hand side of Figure~\ref{fig}.
First, there is a point $\dualPhi=\diag(2m,-2m)$ where one pair of the dual quarks $\dualQ_1,\tilde{\dualQ}^1$
becomes massless. This point actually corresponds to the continuation of the point A. 
In other words,  the dual quarks $\dualQ_1, \tilde{\dualQ}^1$ originate from the magnetic monopoles of the electric theory.
Other points are realized in the region where $\dualPhi$ is small and the dual theory is strongly coupled.
 In this region we can integrate out $\dualQ_1,\tilde{\dualQ}^1$
and get $\dualSU(2)$ theory with $N_f=3$ massless flavors. 
In this $N_f=3$ theory, we have a point B where a dyon of the magnetic theory becomes massless,
and a point C where four monopoles of the magnetic theory become massless. 
So the dyon of the magnetic theory is the dyon of the electric theory,
and the four monopoles of the magnetic theory are the four quarks of the electric theory. 

The situations on the left hand side and on the right hand side of Figure~\ref{fig} are smoothly connected when we change the electric coupling $g$
from small to large values. More details are discussed in Appendix~\ref{sec:Sd}.

\paragraph{Explicit breaking to \Nequals1:}

Now that we reviewed the situation with \Nequals2 supersymmetry, let us turn on supersymmetry breaking terms; in any case we wanted to discuss gapped systems. 
We first turn on the superpotential $W_\text{\Nequals1}=\frac{1}{2}m_\Phi \Tr \Phi^2 = \frac{1}{2}m_\dualPhi \Tr \dualPhi^2$ which breaks \Nequals2  to \Nequals1.
Then the moduli space of $u$ is lifted except for the points A, B and C; 
the low energy effective theory near these points is given as
\beq
W_{\rm eff} = -c (u-u_p) \tilde{E}E +m_\Phi u
\eeq
where $u_p$ ($p=A,B,C$) is the point A, B or C, 
$E, \tilde{E}$ represent the hypermultiplet corresponding to the monopole, dyon or quarks which are charged under the $\U(1)$, and $c$ is a constant.
Then we get a gapped vacuum at $\tilde{E}E=m_\Phi/c$ and $u=u_p$.
In this way, the vacua are now discrete and realized at the points A, B and C where
the quarks, monopoles or dyons condense.

We will  use the vacuum on the point A in our analysis.
The reason is as follows. 
In the electric theory, we want the quarks to have a single mass $m$ so as to realize the $\nu{=}16$ SPT phase.
Then we can use either the point A or B. 
In the dual magnetic theory, the point A
can be analyzed perturbatively, because the gauge group is completely broken by the vevs of $\dualQ_1, \tilde{\dualQ}^1$ and $\dualPhi$ as we will discuss more detail later. 
This makes the vacuum A  easy to analyze in the dual theory.
This is the power of S-duality: the point A is strongly coupled in the original electric theory, but it is weakly coupled by the Higgsing of the gauge group in the dual theory.
This step is dynamically the most non-trivial one in our construction.
The rest of the analysis is technically tedious but is straightforward nonetheless.

\subsection{Continuous deformation}\label{subsec:continuous}
\begin{figure}
\centering
\begin{tabular}{c@{\qquad}c}
\includegraphics[width=.65\textwidth]{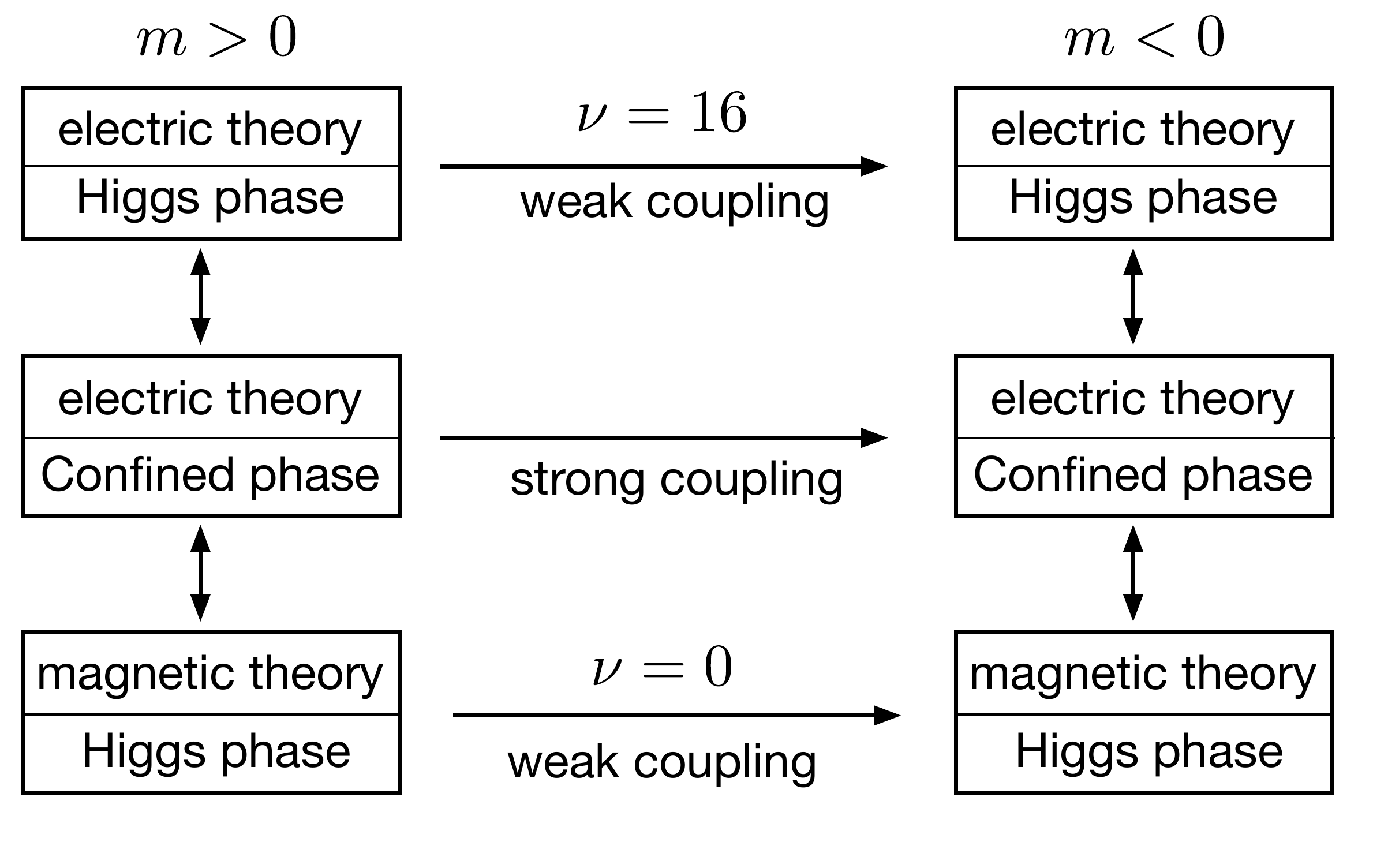}\\
\end{tabular}
\caption{The overall picture. In the electric theory, the Higgs phase and the confined phase can be smoothly connected as we discussed in Sec.~\ref{subsec:general}. 
The confined phase of the electric theory is dual to the Higgs phase of the magnetic theory. When the theory is Higgsed, we can perform weakly coupled analysis.
In the Higgs phase of the electric theory, we get $\nu=16$ when we change the mass $m$ from positive to negative as argued in Sec.~\ref{subsec:general}. 
In Sec.~\ref{subsec:continuous}, 
we are going to show that the Higgs phase of the magnetic theory has $\nu=0$.
\label{fig:picture}}
\end{figure}

After these preparations, we can finally show an explicit continuous deformation from the $\nu{=}16$ phase of the free Majorana fermions to the trivial $\nu{=}0$ phase.
Before going to the technical analysis, we summarize the overall picture in Figure~\ref{fig:picture}.
As in Sec.~\ref{subsec:general}, we start by embedding the $\nu{=}16$ free fermion system into the $\SU(2)$ gauge theory in the Higgs phase.  
We then continuously deform this electric theory to the confined phase of the softly-broken \Nequals2 theory at the point A.
Turning the electric gauge coupling to be very strong,
this can be mapped to  the Higgs phase of the dual magnetic theory.
We will check that this dual magnetic phase has $\nu=0$. 
In this way, our construction explicitly realizes the situation
that in one parameter region of the theory we have $\nu=16$ massless boundary fermions and in another parameter region we get $\nu=0$.
Similar situation in $1+1$ dimensional Gross-Neveu model was discussed in \cite{Fidkowski:2009dba}.

Our analysis of the  continuous deformation consists of seven steps:
\begin{itemize}
\item Step 1: Embedding the $\nu={16}$ theory to the \Nequals2 system,
\item Step 2: Performing the S-duality,
\item Step 3: Pinning the adjoint vev,
\item Step 4: Determination of the vev of the squarks,
\item Step 5: Determination of the dual $\sCP$ transformation,
\item Step 6: Decoupling of unwanted scalars,
\item Step 7: Analysis of the fermions.
\end{itemize}
We will detail each of them in turn below. 
Before proceeding, we note that Step 1 is on the electric side, and
we immediately go to the magnetic side in Step 2. 
The rest of the analysis will be all done on the magnetic side.

\paragraph{Step 1: Embedding the $\nu={16}$ theory to the \Nequals2 system:} 
We start from the electric theory where the electric coupling constant $g$ is very small and
all the superpartners of the $\nu{=}16$ fermions and the $\SU(2)$ gauge field have very large masses. 
This theory reduces to the non-supersymmetric gauge theory
where we just have $\nu{=}16$ Majorana fermions coupled to $\SU(2)$ gauge group.
We discussed at length  in Sec.~\ref{sec:pathint} that  this gauge theory is 
continuously connected to the system of 16 free massive Majorana fermions.
Actually we can recycle some of the scalar fields of the \Nequals2 theory for this purpose as the system $Y(\mu)$ used in Sec.~\ref{subsec:general} by introducing large SUSY breaking potential to them to go from the confined to the Higgs phase.
Now, we continuously make the SUSY breaking terms to be very small, so that the system can be considered as a small deformation of the \Nequals2 system at the vacuum A.

\paragraph{Step 2: Performing the S-duality:} 
We now  deform the electric coupling $g$ to be very large. 
A standard analysis of supersymmetric gauge dynamics on the $u$-plane described in Appendix~\ref{sec:Sd} shows that this process is smooth without any phase transition. 
The dual coupling is given by  $\dualg \sim 1/g  \to 0$, which becomes infinitely weak.
This is the most nontrivial dynamical step of our analysis. 
In the electric theory, the gauge coupling eventually becomes strong by the effect of renormalization group flows
below the mass scale $m$, even if we set the coupling $g$ to be small in the UV. 
However, in the dual magnetic theory, the gauge group is Higgsed.
Thus if the dual UV coupling $\dualg$ is small, all the analysis can be done perturbatively without any strong dynamics.

\paragraph{Step 3: Pinning the adjoint vev:} 
In the following, we pick a positive constant $m_0$, and we change the mass parameter $m$ from $m=m_0>0$ to $m=-m_0<0$.
 We always keep $m_\dualPhi$ to be positive.
To pick the point A as our vacuum when $m=\pm m_0$, we introduce a SUSY breaking potential 
\beq
V_{\CN{=}0}= \lambda | \Tr \dualPhi^2 -8m_0^2|^2 +\lambda' \Tr [\dualPhi, \dualPhi^\dagger]^2
\label{eq:VN0}
\eeq
on the dual side. 
The vev of $\dualPhi$ is then \begin{equation}
\dualPhi\sim \diag(2m_0,-2m_0)\label{eq:adjvev}
\end{equation}
up to very small corrections in $\dualg$. This vev breaks $\dualSU(2)$ to $\dualU(1)$.

We  would like to take $\lambda$ and $\lambda'$ in \eqref{eq:VN0} to be sufficiently large to pin the vev of $\dualPhi$, while keeping everything perturbative.
Recall that we are using a convention common in \Nequals2 studies such that the kinetic term for $\dualPhi$ is non-canonical, 
such that the K\"ahler potential is given by $K = 1/\dualg^2 \Tr (\dualPhi^\dagger \dualPhi)$. 
After canonically normalizing the fields,
the condition of perturbativity is given by $\dualg^4\lambda, \dualg^4\lambda' \ll 1$. 
To pin  the vacuum at \eqref{eq:adjvev}
we need
 $\lambda m^4_0 \gg \dualg^2 m^2_\dualPhi m^2_0$ so that the contribution of $V_{\CN{=}0}$ to the total potential is much more significant than the SUSY preserving ones which
 include a term $\dualg^2 |m_\dualPhi \dualPhi +\cdots|^2$. 
Later we need to impose the condition $\dualg^2 m_\dualPhi > 2m_0$ in \eqref{eq:higgscondition}, and hence we need $\lambda \gg \dualg^{-2}$.
In summary, we take $\lambda, \lambda'$ to be in the region $\dualg^{2} \ll \dualg^{4}\lambda, \dualg^{4}\lambda' \ll 1$. 
For simplicity we consider the formal limit in which $\lambda, \lambda' \to \infty$, $\dualg\to0$ such that the condition just stated is satisfied.
Now, the system  has the minimum at the point A when $m=\pm m_0$.

When $m$ deviates from $\pm m_0$, the point of massless dual quark on the $u$-plane at $\dualPhi=(2m,-2m)$ is different from the point \eqref{eq:adjvev}.
We always keep \eqref{eq:adjvev} by the above potential.

\paragraph{Step 4: Determination of the vev of the squarks:}
At this point the potential for the scalar components of the dual quarks $\dualQ$, $\tilde\dualQ$ is still given by that of the \Nequals2 theory. 
Using the standard formula for the supersymmetric Lagrangian and replacing $\dualPhi$ by the vev given above, the potential of the scalar components of the dual quarks is now given by
\beq
V=V_F+V_D,\label{eq:pot}
\eeq
where $V_F$ is obtained from the superpotential $W= \sum_{i=1}^{4} (-\tilde{\dualQ}^i \dualPhi \dualQ_i ) +   2m \tilde{\dualQ}^1  \dualQ_1+ \frac{1}{2}m_\dualPhi \Tr \dualPhi^2$ as
\beq
&V_F = \frac{\dualg^2 }{2} \left| 4 m_\dualPhi m_0  - \sum_{i=1}^4 (\dualQ^+_i \tilde{\dualQ}_+^i - \dualQ^-_i \tilde{\dualQ}_-^i) \right|^2 
+\dualg^2  \left| \sum_{i=1}^4\dualQ^+_i \tilde{\dualQ}_-^i  \right|^2+\dualg^2  \left| \sum_{i=1}^4\dualQ^-_i \tilde{\dualQ}_+^i  \right|^2  \nonumber  \\
& +4(m_0-m)^2( |\dualQ^+_1|^2+|\tilde{\dualQ}_+^1|^2)+ 4(m_0+m)^2 ( |\dualQ^-_1|^2 + |\tilde{\dualQ}_-^1|^2)  \nonumber \\
&+4m_0^2 \sum_{j=2}^4 (|\dualQ^+_j|^2+|\dualQ^-_j|^2+|\tilde{\dualQ}_+^j|^2+|\tilde{\dualQ}_-^j|^2), 
\eeq
and $V_D$ is the $D$-term potential given by
\beq
V_D  = \frac{\dualg ^2}{8}  \left(\sum_{i=1}^4  (  | \dualQ^+_i |^2 - | \dualQ^-_i |^2- |\tilde{\dualQ}^i_+|^2 +  |\tilde{\dualQ}^i_-|^2 ) \right)^2 
+ \frac{\dualg ^2}{2} \left| \sum_{i=1}^4  (  \dualQ^+_i  \overline{\dualQ^-_i} - \tilde{\dualQ}^i_- \overline{\tilde{\dualQ}^i_+} ) \right|^2 .
\eeq
where $\pm$ on the quark fields are the $\dualSU(2)$ indices $a=\pm$, or more explicitly we are using the notations $\dualQ_i=(\dualQ^a_i)=(\dualQ^+_i, \dualQ^-_i)^T$ and 
$\tilde{\dualQ}^i=(\tilde{\dualQ}_a^i)=(\tilde{\dualQ}_+^i, \tilde{\dualQ}_-^i)$.


When $m=\pm m_0$, we see that the vacuum is given by 
\beq
\dualQ_1&=(2 \sqrt{m_\dualPhi m_0}, 0)^T,&\tilde{\dualQ}^1&=(2\sqrt{m_\dualPhi m_0}, 0), & (m&=m_0>0), \label{eq:qvev1} \\
\dualQ_1&=(0, 2 \sqrt{m_\dualPhi m_0})^T,&\tilde{\dualQ}^1&=(0, -2\sqrt{m_\dualPhi m_0}), &(m&=-m_0<0).\label{eq:qvev2}
\eeq
The vevs of $\dualQ_i$ and $\tilde{\dualQ}^i$ for $i=2,3,4$ are all zero. 

\paragraph{Step 5: Determination of the dual $\sCP$ transformation:}
At this point, we determined the vevs of all the fields on the dual side, and 
we can complete the determination of the $\sCP$ transformation on the dual side,
whose last step was left unfinished at the end of Sec.~\ref{subsec:dualCP}.

The action of $\sCP$ on the dual adjoint field is simply given by \begin{equation}
\sCP(\dualPhi)= \dualPhi^\dagger
\end{equation}
and then the vev $\dualPhi=\diag(2m,-2m)$ \eqref{eq:adjvev} is invariant.

On the dual quarks, the action can be written as  follows. 
First, we define $\sCP_0$ which commutes with the $\dualSU(2)$ gauge group as
\beq
\sCP_0 :=\sC_{\SU(2)R} \sC_{\SO(8)} \sPinv.
\eeq
However, this $\sCP_0$ has two problems. First, its square is given by $(\sCP_0)^2=(-1)^F(-1)^G$, where $(-1)^G$ is the center of $\dualSU(2)$; $(-1)^G=-1$ for 
dual quarks and $(-1)^G=+1$ for other fields. But we want the relation $\sCP^2=(-1)^F$ so that the dual theory can be put on a $\pin^+$ manifold.
Second, this $\sCP_0$ is broken by the vevs of the dual quarks given above, because its action on $\dualQ_1$ and $\tilde{\dualQ}^1$ is given by 
$\sCP_0(\dualQ_1)=\ii \overline{\tilde \dualQ^1}$ and $\sCP_0(\tilde{\dualQ}^1)=-\ii \overline{\dualQ_1}$. 

These problems can be solved at the same time by
introducing a gauge transformation $\sC_{\dualSU(2)} \in \dualSU(2) $ given by
\beq
\sC_{\dualSU(2)} = & \diag(-\ii,\ii) 
\eeq
and then defining
\beq
\sCP:= & \sC_{\dualSU(2)} \sCP_0= \sC_{\dualSU(2)} \sC_{\SU(2)R} \sC_{\SO(8)} \sPinv. \label{eq:modifiedCP}
\eeq
Under this $\sCP$, the transformation of the fields are given as
\beq
\sCP (\dualQ_1^\pm)  &= \pm \overline{\tilde{\dualQ}^1_\pm}, &
\sCP (\tilde{\dualQ}^1_\pm)  &= \pm \overline{{\dualQ}_1^\pm}, 
 \nonumber \\
\sCP (\dualQ_i^\pm) &= \mp \overline{\tilde{\dualQ}^i_\pm},  &
\sCP (\tilde{\dualQ}^i_\pm)  &= \mp \overline{{\dualQ}_i^\pm}, &
 (i&=2,3,4). \label{eq:dualCP}
\eeq
They preserve the vevs \eqref{eq:qvev1}, \eqref{eq:qvev2} of the scalar components of the dual quarks.

The fact that we need to mix the gauge transformation $\sC_{\dualSU(2)}$ to the definition of $\sCP$ implies that the symmetry group
of the dual theory is not simply $\dualSU(2) \times \Pin^+$ but more complicated. We will discuss more on this point later in Sec.~\ref{subsec:pin}.

\paragraph{Step 6: Decoupling of unwanted scalars:}
When $m=\pm m_0$, the only nonzero vev of the scalar components of the dual quarks are $\sCP$-even,
and the vev is neutral under $\SO(6)$ which act on $\dualQ_i$, $\tilde{\dualQ}^i$ for $i=2,3,4$.
The potential $V$ \eqref{eq:pot} is at most quartic, and invariant under $\sCP$ and $\SO(6)$.
We can then safely add large mass terms to the scalars charged under $\SO(6)$,
and to the scalars neutral under $\SO(6)$ but is odd under $\sCP$, to remove them.

More concretely, we proceed as follows. First we add mass terms,
\beq
V_{\rm add1} =M^2\sum_{i=2}^4(| \dualQ_i |^2 + | \tilde{\dualQ}^i|^2).
\eeq
As long as $M$ is much larger than other mass scales, we can integrate out the $\SO(6)$ charged quarks $\dualQ_i$, $\tilde{\dualQ}^i$ for $i=2,3,4$ and set them to be zero.
Next, we add a gauge and $\sCP$ invariant potential
\beq
V_{\rm add2} &=\lambda''  \left| (2m_0+\dualPhi ) \dualQ_1 -(2m_0+\dualPhi^\dagger ) (\tilde{\dualQ}^1)^\dagger \right|^2
+\lambda''  \left| (2m_0-\dualPhi ) \dualQ_1 +(2m_0-\dualPhi^\dagger ) (\tilde{\dualQ}^1)^\dagger \right|^2 \nonumber \\
&=16\lambda'' m_0^2 \left( | \dualQ_1^+ -\overline{\tilde{\dualQ}_+^1} |^2+| \dualQ_1^- +\overline{\tilde{\dualQ}_-^1} |^2 \right),
\eeq
where we set $\dualPhi=\diag(2m_0, -2m_0)$.
This gives masses to the $\sCP$ odd scalars $\dualQ_1^+ -\overline{\tilde{\dualQ}_+^1}$ and $ \dualQ_1^- +\overline{\tilde{\dualQ}_-^1}$
and we can set them to zero if $\lambda''$ is large enough.

The remaining scalars are complex fields $z$ and $w$ given by 
\beq
z:=\dualQ^+_1 =\overline{\tilde{\dualQ}_+^1}~~~w:=\overline{\dualQ_1^-}=-\tilde{\dualQ}_-^1.
\eeq
Then, the potential \eqref{eq:pot} simplifies to
\beq
V =\frac{\dualg^2 }{2} \left| 4 m_\dualPhi m_0  - |z|^2-|w|^2 \right|^2 
+4\dualg^2  \left| zw  \right|^2  +8(m_0-m)^2 |z|^2+ 8(m_0+m)^2  |w|^2 .
\eeq
This is now far easier to analyze. 

Now we can find the potential minimum for general $m$. 
We require that parameters satisfy the relation
\beq
\dualg ^2 m_\dualPhi > 2m_0. \label{eq:higgscondition}
\eeq
Under this condition, when $m \neq 0$, the minimum is given by
\beq
|z|^2 &= \frac{4\dualg ^2m_\dualPhi m_0 - 8(m_0-m)^2}{\dualg ^2},&w&=0, & (m&>0),\\
|w|^2 &= \frac{4\dualg ^2m_\dualPhi m_0 - 8(m_0+m)^2}{\dualg ^2},&z&=0,& (m&<0). 
\eeq
The phase of $z$ or $w$ is eaten by the $\dualU(1)$ gauge field by the Higgs mechanism and they become massive together.
There is a unique vacuum with a mass gap.

When $m=0$, the minimum of $V$ is realized by $(z,w)$ satisfying the conditions $|z|^2+|w|^2 = (4\dualg ^2m_\dualPhi m_0 - 8m_0{}^2)/\dualg^2 $ and $ zw=0$. 
This potential itself leads to a first order phase transition from $(z \neq 0, w=0)$ to $(z =0, w \neq 0)$. 
To avoid such a phase transition,
we further deform the potential by adding $\dualSU(2)$ and $\sCP$ symmetric terms given by 
\beq
V_{\rm add3} & = -\frac{1}{4} \dualg'^2 | \bar{\dualQ}^1 \tau^A \dualQ_1 - \tilde{\dualQ}^1 \tau^A \bar{\tilde{\dualQ}}_1  |^2=-4\dualg'^2 | z w|^2, \\
V_{\rm add4} &= \lambda''' | \mu -  \tilde{\dualQ}^1\dualQ_1  |^2 = \lambda''' \left( \mu -|z|^2+|w|^2 \right)^2, \\
V_{\rm add5} & = -\epsilon( \bar{\tilde{\dualQ}}_1 J \dualQ_1 +{\rm c.c.} )= -2\epsilon (z\bar{w}+{\rm c.c.}).
\eeq
Let us explain the roles of each of them. 
\begin{itemize}
\item The $V_{\rm add3}$ term cancels the term $4\dualg^2 |zw|^2$ by taking $\dualg' \to \dualg $, which makes the analysis of the potential a little easier (but this is not absolutely necessary).
After turning on $V_{\rm add3}$, the potential minima are given by $|z|^2+|w|^2=(4\dualg^2  m_\dualPhi m_0 -8m_0^2)/\dualg^2 $.
\item Then by turning on $V_{\rm add4}$, we can fix the ratio of the absolute values of $z$ and $w$, $|z|/|w|$, to whatever values we want.
Thus we can smoothly connect the points $(z \neq 0, w=0)$ to $(z=0, w \neq 0)$ by smoothly changing the parameter $\mu$.
\item Finally, in the intermediate region where both $z$ and $w$ are nonzero, there remains a massless boson coming from the relative phase $\arg (z \bar{w})$.
This is the Goldstone boson associated to the $\U(1)$ flavor symmetry acting on $(\dualQ_1, \tilde{\dualQ}^1)$.
(The overall phase is absorbed by the dual $\dualU(1)$ gauge field.) By turning on $V_{\rm add5}$ with small $\epsilon$, this Goldstone boson is eliminated since
this $V_{\rm add5}$ breaks the flavor $\U(1)$ symmetry.
\end{itemize}
This completes the argument that we can continuously deform from $m=m_0>0$ to $m=-m_0<0$
without having massless bosons (i.e., scalars or gauge fields). In particular, the vevs of $z$ and $w$ do not break $\sCP$,
so the $\sCP$ is preserved during this continuous deformation.

Let us note the following point, in relation to the criterion we found in Sec.~\ref{subsec:flavor}.
Before adding the term $V_{\rm add5}$, the theory has the flavor symmetry $\U(4)$ which is preserved by the mass $m$.
After adding the $V_{\rm add5}$, the flavor symmetry is reduced to $\SU(4)$ which is the double cover of $\SO(6)$. 
Here we consider the double cover $\SU(4)$ instead of $\SO(6)$ because the quarks of the original electric theory are in the fundamental representation of $\SU(4)$.
Then we can avoid the no-go argument given in section~\ref{subsec:flavor}
because this $\SU(4)$ is simple, connected and simply connected and has $t_F=2$ in the notation of that section. Note that the value $t_F=2$
is realized in different ways in the electric theory and magnetic theory. In the electric theory, we have 2 copies of the
${\bf 4}+\bar{\bf 4}$ dimensional representation of $\SU(4)$.
In the magnetic theory, we have 2 copies of the ${\bf 6}$ dimensional representation of $\SU(4)$ which is the vector of $\SO(6)$. 
In fact, if $\U(4)$ were preserved, in particular we would have $(\bZ_2)^4 \subset \U(4)$ where each $\bZ_2$ acts on each flavor of quarks $Q_i, \tilde{Q}^i$ in the electric theory.
Then by completely the same argument as the one around \eqref{eq:etaincludeflavor}, it would be impossible to make the boundary theory trivial.
Thus the term $V_{\rm add5}$ is really crucial.

\paragraph{Step 7: Analysis of the fermions:}
Now let us consider the fermion masses. 
In the process of continuous deformation,
the scalar components of the dual quarks $\dualQ_j, \tilde{\dualQ}^j~(j=2,3,4)$ never get a vev. Then, the fermions contained in $\dualQ_j, \tilde{\dualQ}^j~(j=2,3,4)$ do not mix with other fermions and their masses
are given by the vev $\dualPhi=(2m_0, -2m_0)$ which is constant. Therefore these fermions do not become massless during the deformation.
The remaining fermions are $3+3+2+2=10$ fermions coming from the \Nequals1  gauginos $(\lambda^{++},\lambda^{--}, \lambda^0)$, 
the fermions in $\dualPhi$ ($\psi^{++},\psi^{--}, \psi^0$) and in $\dualQ_1, \tilde{\dualQ}^1$ $(\frakq^+,\frakq^-, \tilde{\frakq}_+,\tilde{\frakq}_-)$.
This implies that we can realize at most the $|\nu| \leq 10$ SPT phase. 
Now, note that $\nu \bmod 16$ is preserved by coupling to the gauge fields 
as we discussed in section~\ref{sec:pathint} because the partition function depends on $\nu \bmod 16$. Thus the only logical possibility is $\nu{=}0$.
This simple argument guarantees our success.
For those who are not satisfied by the above argument, the fermion mass matrix is treated explicitly in Appendix~\ref{sec:mass}.

\subsection{How the argument fails for $\nu{=}8$}\label{subsec:nu8}
Up to now, we have taken the masses of quarks $Q_i, \tilde{Q}_i$ as $m_1=m_2=m_3=m_4=m$.
However, we can also consider other masses such as
\beq
m_1=m_2=m_a,~~~m_3=m_4=m_b.
\eeq
Then we can change only $m_b$ from $m_0$ to $-m_0$ while fixing $m_a$ at $m_a=m_0$. 
This case corresponds to the case of the $\nu{=}8$ SPT phase because we are only changing 
the masses of half the quarks. In this case, we should not be able to deform the parameters while preserving the mass gap and without breaking the $\sCP$.
Let us look at what happens.

The dual quark masses are given by
\beq
\dualm_1=m_a+m_b,~~\dualm_2=m_a-m_b,~~\dualm_3=\dualm_4=0.
\eeq
Let us change $m_b$ from positive to negative values while fixing $m_a$. When $m_b>0$,
the potential minimum is realized by giving a vev to $\tilde{\dualQ}_+^1 , \dualQ^+_1$. When $m_b<0$,
the potential minimum is realized by a vev of $\tilde{\dualQ}_+^2 ,\dualQ^+_2$. 

However, recall that the action of $\sCP$ was given as in \eqref{eq:dualCP}.
From this, one can check that 
the vev of $\tilde{\dualQ}_+^1 ,\dualQ^+_1$ \emph{does} preserve the $\sCP$ in the region $m_b>0$ but that
the vev of $\tilde{\dualQ}_+^2 ,\dualQ^+_2$ \emph{does not} preserve the $\sCP$ in the region $m_b<0$.
This originates from the difference of the actions of $\sC_{\dualSO(8)}$ on $\tilde{\dualQ}_+^1 ,\dualQ^+_1$ and $\tilde{\dualQ}_+^j ,\dualQ^+_j~(j=2,3,4)$.
This does not mean that we do not have any $\sCP$ symmetry in the region $m_b<0$.
We can take a gauge transformation $(-1)^G:=\diag(-1,-1) \in \dualSU(2)$ and define a new $\sCP$ as
\beq
\sCP'=(-1)^G\sCP. \label{eq:anotherCP}
\eeq
Notice that this new $\sCP'$ also satisfies $\sCP'^2=(-1)^F$.
However, the definition of the unbroken $\sCP$ changes when we pass through the region $m_b=0$.

Let us consider a deformation process from $m_b=m_0$ to $m_b=-m_0$. We assume that the adjoint scalar vev $\dualPhi=\diag(2m_0,-2m_0)$ is fixed,
but the vevs of other scalars can be arbitrary by introducing various terms in the potential. 
We can mix gauge transformations to $\sCP$, but to satisfy $\sCP^2=(-1)^F$ and keep the adjoint vev $\dualPhi$ invariant, 
the only possible $\sCP$ are \eqref{eq:modifiedCP} and \eqref{eq:anotherCP}.
We need to consider two cases separately:
\begin{itemize}
\item During the continuous deformation, we go through a parameter region where both $\sCP$ and $\sCP'=(-1)^G \sCP$ are broken.
\item During the continuous deformation, we go through a parameter region where both $\sCP$ and $\sCP'=(-1)^G \sCP$ are unbroken.
\end{itemize} 
In the first case, $\sCP$ is broken in that parameter region and we cannot smoothly connect $m_b=m_0$ and $m_b=-m_0$.
In the second case, the $(-1)^G \in \dualSU(2)$ is unbroken in the parameter region where both $\sCP$ and $\sCP'=(-1)^G \sCP$ are unbroken.
However, the $(-1)^G$ is broken at the initial and final points of the deformation $m_b=m_0$ and $m_b=-m_0$, 
and hence we have to encounter a phase transition during the continuous
deformation at which $(-1)^G$ is recovered.

Therefore, in either case, we \emph{cannot} show that the SPT phases corresponding to $m_b=m_0$ and $m_b=-m_0$ are the same. 
Of course this is as expected since $\nu=8$ is guaranteed to be nontrivial by the consideration of the phase of the partition function on $\RP^4$, but it is reassuring that we also obtain  a consistent result from this analysis.

\subsection{Pin structure in the dual theory}\label{subsec:pin}
We have  established that the $\nu{=}16$ phase and the $\nu{=}0$ phase can be continuously connected on a flat space, as schematically the situation shown in Figure~\ref{fig:picture}.
During the discussion, we have found that we have to mix a gauge transformation $\sC_{\dualSU(2)} =  \diag(-\ii,\ii) $ to define the $\sCP$ action on the dual magnetic theory, as
\beq
\sCP= & \sC_{\dualSU(2)} \sCP_0= \sC_{\dualSU(2)} \sC_{\SU(2)R} \sC_{\SO(8)} \sPinv. \label{eq:mixedgaugeCP}
\eeq

Let us consider putting the theory on an unoriented $\pin^+$ manifold. Then the above fact means that we have to mix gauge transformations 
in the definition of the $\pin^+$ structure. Let us see more precisely how it works.
The reason we discuss it here is to make sure that there is no inconsistency
analogous to the anomaly of the spin-charge relation discussed in \cite{Seiberg:2016rsg}.

For this purpose, we have to identify the symmetry group of the theory which is left unbroken by various mass and deformation parameters.
The $\sCP_0$ commutes with the gauge symmetry, but its square is $(\sCP_0)^2=(-1)^F(-1)^G$ and 
hence the unbroken group cannot be simply $\Pin^+(3,1) \times \dualSU(2)$.
It also implies that if we naively forget the gauge group $\dualSU(2)$, the theory cannot be put on a $\pin^+$ manifold
because the quarks would have the relation $(\sCP_0)^2=1$, meaning that we need a  $\pin^-$ structure,
but a $\pin^-$ structure cannot be put on a generic $\pin^+$ manifold.\footnote{A quick way to see this is as follows.
Let $M$ be a $d$-dimensional manifold and $\CE=\wedge^d TM$ be its orientation line bundle. 
Let $E_{d+1}=TM \oplus \CE$ and $F_{d+3}=TM \oplus \CE \oplus \CE \oplus \CE$ be orientable bundles.
Then the spin structures of $E_{d+1}$ and $F_{d+3}$ reduce to 
the $\pin^-$ and $\pin^+$ structures of $M$, respectively. 
The Stiefel-Whitney class of $TM$ is denoted as $w=1+w_1+w_2+\cdots$.
Then the Stiefel-Whitney classes of $E_{d+1}$ and $F_{d+3}$ are given by
$w(E_{d+1})=(1+w_1+w_2+\cdots)(1+w_1)=1+(w_2+w_1^2)+\cdots$ and $w(F_{d+3})=(1+w_1+w_2+\cdots)(1+w_1)^3=1+w_2+\cdots$,
where we have used the fact that the line bundle $\CE$ has the Stiefel-Whitney class $1+w_1$. Therefore, a $\pin^-$ structure requires $w_2(E_{d+1})=w_2+w_1^2=0$,
while a $\pin^+$ structure requires $w_2(F_{d+3})=w_2=0$. This means that a manifold with $w_1^2 \neq 0$ can have at most only one of $\pin^-$ or $\pin^+$, but not both.
For example, the Stiefel-Whitney class of $\mathbb{RP}^d$ is given by $w=(1+a)^{d+1}$, 
where $a$ is the generator of $H^1(\mathbb{RP}^d, \bZ_2)$. Thus $w_2(E_{d+1})=\frac{1}{2}(d+1)(d+2)  a^2$ and $w_2(F_{d+3})=\frac{1}{2}d(d+1)  a^2$.
If we put $d=4$, we can see that $\mathbb{RP}^4$ has $\pin^+$ but it cannot have $\pin^-$.}
So let us see what is going on.

Before introducing various deformations, the dual $\CN=2$ theory with $\CN=2$ preserving mass term has the symmetry group 
$\dualSU(2) \times \SU(2)_R \times \SO(2) \times \SO(6) \times \Pin^-(3,1)$ at the level of its Lagrangian.
Here $\Pin^-(3,1)$ is defined by using $\sPinv$ which came from the $5=4+1$ dimensional Lagrangian. 
Let $(-1)^G \in \dualSU(2)$, $(-1)^R \in \SU(2)_R$ and $(-1)^F \in \Pin^-(3,1)$ be the centers of the respective groups.
One can check that the product $(-1)^G(-1)^R(-1)^F$ acts trivially on all the fields of the theory, so the symmetry group is actually
\beq
\frac{\dualSU(2) \times \SU(2)_R \times \Pin{}^- (3,1)}{\bZ_2 }\times \SO(2) \times \SO(6)
\eeq
where $\bZ_2$ is generated by $(-1)^G(-1)^R(-1)^F$.

After introducing deformations, this is broken to a subgroup. To describe it, take $(\bZ_{4})_R \subset \SU(2)_R$ which is a $\bZ_4$ subgroup generated by $\sC_{\SU(2)R}$; $(-1)^R$ is a unique order-2 element in this $(\bZ_{4})_R$.
We define a homomorphism $\pi_1:\Pin^- \to \bZ_2$ such that elements of $\Pin{}^- $ which reverse orientation map to the element $(-1) \in \bZ_2$,
and also define $\pi_2:(\bZ_{4})_R\to \bZ_2 $ such that $\pi_2(\sC_{\SU(2)R})=(-1)$. Furthermore, we take a $\bZ_2$ subgroup $(\bZ_{2})_S \subset \SO(2)$.
Then we define a new group $\tilde{\Pin}(3,1)$ as
\beq
\tilde{\Pin}(3,1):=\{ (c, d, e) \in(\bZ_{4})_R \times (\bZ_2)_S \times \Pin{}^-  ;\  \pi_2(c)=d=\pi_1(e)   \}.
\eeq
This contains the elements $(-1)^R$, $(-1)^F$ and $\sCP_0=\sC_{\SU(2)R} \sC_{\SO(8)} \sPinv$, and
it is a double cover of both $\Pin^-(3,1)$ and $\Pin^+(3,1)$.
Indeed, the projection to the third component just gives $\Pin^-(3,1)$ which is $2:1$,
and the $\bZ_2$ quotient with respect to $(-1)^R(-1)^F$ is $\Pin^+(3,1)$.

Then the unbroken group after the deformations is
\beq
\dualSU(2) \times_{\bZ_2}  \tilde{\Pin}(3,1) := \frac{\dualSU(2) \times  \tilde{\Pin}(3,1) }{\bZ_2 } \label{eq:magstructuregroup}
\eeq
times $\SO(6)$ which is uninteresting to us.
Here the $\bZ_2$ quotient is taken with respect to the element $(-1)^G(-1)^R(-1)^F$ as before.
This group \eqref{eq:magstructuregroup} is the structure group of the magnetic theory when we put the theory on a nontrivial manifold.
This is an analog of the $\Pin^c$ group.

There is a homomorphism from the group $\dualSU(2) \times_{\bZ_2}  \tilde{\Pin}(3,1)$ to $\Pin^+(3,1)$ such that the following diagram commutes:
\beq
\xymatrix{
\dualSU(2) \times_{\bZ_2}  \tilde{\Pin}(3,1)  \ar[rr] \ar[rrd] & &\Pin{}^+(3,1)   \ar[d] \\
& & \OO(3,1).
}\label{eq:hom}
\eeq
This can be described as follows. Mathematically, we have a forgetful map from the left hand side to $\tilde{\Pin}(3,1)/\bZ_2$, which equals $\Pin^+(3,1)$ as discussed above. 
More physically, consider a gauge invariant fermion operator $\dualPsi$ of the theory. For example, we may take
\beq
\dualPsi=\left(
\begin{array}{c}
\Tr \dualPhi^\dagger \lambda_\alpha \\
\Tr \dualPhi \bar{\lambda}^\aa 
\end{array}
\right).
\eeq 
where $\lambda$ is the gaugino.
Then the transformations in $\dualSU(2) \times_{\bZ_2}  \tilde{\Pin}(3,1)$
acts as transformations in $\Pin^+(3,1)$ on $\dualPsi$.
Therefore, by looking at the transformation of $\dualPsi$, we can define the above homomorphism.

If we are given a $\pin^+$ manifold $M$, the theory can be put on $M$ as follows. 
The $\pin^+$ bundle is defined on $M$ by a set of transition functions. We
specify an uplift of transition functions from $\Pin^+(3,1)$ to $ \dualSU(2) \times_{\bZ_2}  \tilde{\Pin}(3,1)$,
\beq
\Pin{}^+(3,1)  \to \dualSU(2) \times_{\bZ_2}  \tilde{\Pin}(3,1) 
\eeq
such that it is consistent with the homomorphism \eqref{eq:hom}.
There exists such uplift of the bundle. 
For example we can use \eqref{eq:mixedgaugeCP} for this purpose.
For the existence of this uplift, the division by $\bZ_2=\{1, (-1)^G(-1)^R(-1)^F \}$ in \eqref{eq:magstructuregroup} is important, because the square of $\sCP$ is given by
\beq
\sCP^2= \sC_{\dualSU(2)}^2 \sC_{\SU(2)R}^2 \sPinv^2=(-1)^G(-1)^R
\eeq
where we have used $\sPinv^2=1$. This is equal to $(-1)^F$ only if we impose $(-1)^G(-1)^R(-1)^F=1$.
One consequence of the division by $\bZ_2$ is that if there is a field which is a singlet under $\tilde{\Pin}(3,1)$, then that field must have gauge charge under
$\dualSU(2)/\bZ_2 \simeq \mathitsf{SO}(3)$. 
In particular,  the $\dualSU(2)$ bundle does not exist in itself in general.

Note that the uplift with the above property is not unique, 
but remember that $\dualSU(2)$ is a dynamical gauge group and hence we integrate over all the possible uplifts in the path integral. 
In this way we can consistently define the magnetic theory on a $\pin^+$ manifold.

\section{Conclusions}\label{sec:conclusions}
In this paper, we initiated the study of the effects of strongly-coupled gauge interactions to the topological phases of matter.
We  mainly studied those SPT phases protected by the $\sCP$ symmetry with $\sCP^2=(-1)^F$, in $3{+}1$ spacetime dimensions with relativistic symmetry.

We first discussed under which conditions the introduction of (possibly strongly-coupled) gauge interactions preserve the topological phases. The rule of thumb we found is that if the $\pi_0$ and $\pi_1$ of the gauge group are trivial and the effective theta angle is zero, the system with dynamical gauge field can be continuously connected with the system before the addition of the gauge field, and thus is in the same topological phase.
We gave a general derivation of this statement, and then verified it in more detail in the case of Majorana fermion systems coupled to the gauge field.
We then tested our statement by studying  non-supersymmetric QCD with various groups in the infrared, using the WZW action of the pseudo-Goldstone modes.

Next, we showed that the knowledge of the strong-coupling dynamics allows us to directly show that the $\nu{=}16$ phase of the topological superconductor can be continuously connected to the $\nu{=}0$ phase, thus explicitly demonstrating the collapse of the free fermion classification $\bZ$ to $\bZ_{16}$ due to the interaction effects.
The crucial input we used was the S-duality of \Nequals2 $\SU(2)$ gauge theory with $N_f=4$ flavors. 

Clearly, what was given in this paper is just a tip of the iceberg of the connection between the rich field of topological phases of matter and of the strongly-coupled gauge dynamics. 
A couple of further directions that immediately come to our mind:
\begin{itemize}
\item In this paper, we mainly discussed the gauged topological phases from the bulk point of view, and did not discuss the dynamics of the boundary theory much. We should definitely study them; it might or might not be helpful in finding the boundary theory explicitly. 
\item In this paper, we studied the behavior of the $\sCP$ invariance under duality only in the case of \Nequals2 $\SU(2)$ theory with $N_f=4$ flavors to the minimal extent necessary for our analysis. The result was much subtler than naively expected. It would be interesting to carry out a systematic analysis of the $\sCP$ actions for other known dualities, and to see if they tell us anything new about topological phases of matter.
\end{itemize}
We would like to come back to some of these questions in the future.

\section*{Acknowledgements}
The authors thank Yu Nakayama for a collaboration during the initial stage of this work. 
It was his journal club on the so-called Kitaev-Wen mechanism \cite{BenTov:2015gra,Wen:2013ppa,Wang:2013yta} that started this work, and his contribution was absolutely essential.
The authors also thank Ayuki Kamada for discussions on Wess-Zumino-Witten term.
The work of YT is partially supported in part by JSPS Grant-in-Aid for Scientific Research No. 25870159,
and  by WPI Initiative, MEXT, Japan at IPMU, the University of Tokyo.
The work of K.Y is supported by World Premier International Research Center Initiative
(WPI Initiative), MEXT, Japan.

\appendix
\addtocontents{toc}{\protect\setcounter{tocdepth}{1}}

\section{Basics of $\sCP$ and $\sT$ transformations}\label{sec:CPTbasics}
In this Appendix we review the basics of $\sCP$ and $\sT$ transformations. 
The discussions in this section are mainly to set up the notations and the conventions.

Our conventions for the spinor fields follow those of Wess and Bagger \cite{WessBagger}.
Let us very briefly recall them.
The double cover of the Lorentz group $\SO(1,3)$ is given by $\SL(2,\bC)$, and hence representations of operators and fields under the Lorentz group can be specified by
putting spinor indices of $\SL(2,\bC)$ like $\alpha=1,2$ and $\aa=1,2$ in the fundamental and anti-fundamental representations of $\SL(2,\bC)$, respectively.
Using this notation, a general operator can be represented as $O_{\a_1 \cdots \a_p \bb_1 \cdots \bb_q}$.
Under complex conjugation, an index without dot becomes an index with dot and vice versa: $(O_{\a_1 \cdots \a_p \bb_1 \cdots \bb_q})^\dagger=\overline{O}_{\aa_1 \cdots \aa_p \b_1 \cdots \b_q}$.
These spinor indices are related to the space-time index $\mu=0,1,2,3$ by using four $2 \times 2$ matrices
$(\bar{\sigma}^{ \aa \alpha})^\mu$ given as
\beq
\bar{\sigma}^0=-1,~~~\bar{\sigma}^i=-\tau^i~~~(i=1,2,3)
\eeq
where $\tau^i$ are the standard Pauli matrices.
There are also totally antisymmetric invariant tensors of $\SL(2,\bC)$ denoted as $\epsilon^{\alpha \beta}$, $\epsilon_{\alpha \beta}$, $\epsilon^{\aa\bb}$ and 
$\epsilon_{\aa\bb}$ which are explicitly given as $\epsilon^{12}=-\epsilon^{21}=+1$ and $\epsilon_{12}=-\epsilon_{21}=-1$.
By using them, we can raise and lower indices, e.g., if we have an operator $\psi_\alpha$, 
we define $\psi^\alpha=\epsilon^{\alpha\beta}\psi_\beta$. The $\epsilon_{\alpha \beta}$ is the inverse matrix to $\epsilon^{\alpha \beta}$ 
and hence we get $\psi_\alpha=\epsilon_{\alpha\beta}\psi^\beta$. 
We also use abbreviation; if we have e.g. $\eta_\alpha$ and $\xi_\alpha$, then $\xi \eta :=\xi^\alpha \eta_\alpha$, $\bar{\eta}\bar{\xi} :=\bar{\eta}_\aa \bar{\xi}^\aa$
and $\bar{\xi} \bar{\sigma}^\mu \eta=\bar{\xi}_\aa (\bar{\sigma}^{\aa \a})^\mu \eta_\a$.
More details can be found in Appendix~A of Wess and Bagger \cite{WessBagger}.

\subsection{$\sCPT$ in four dimensions}

Any relativistic unitary theory is symmetric under the $\sCPT$ transformation, and it is given in 4d as follows. 
For arbitrary operator $O_{\a_1 \cdots \a_p \bb_1 \cdots \bb_q}$ it is defined as\beq
\sCPT( O_{\a_1 \cdots \a_p \bb_1 \cdots \bb_q}(x) )= (-1)^q \cdot  \ii^{(p+q)^2} \cdot (O_{\a_1 \cdots \a_p \bb_1 \cdots \bb_q})^\dagger  (-x).\label{eq:defCPT}
\eeq
For a derivation in the axiomatic framework, see e.g.~Sec.~II.5 of \cite{Haag}.
Note that $\sCPT^2=(-1)^F$.

Using $\sCPT$ invariance, $\sT$ and $\sCP$ are related and we can use either of them depending on our preference. We will use $\sCP$ for definiteness. 
Here we note that \begin{equation}
\sT^2=1 \leftrightarrow \sCP^2=1,\qquad
\sT^2=(-1)^F \leftrightarrow \sCP^2=(-1)^F.
\end{equation}
This follows from the fact that $\sT (\sCPT)=(-1)^F (\sCPT)\sT$ because of the factor of $\ii^{(p+q)^2}$ in \eqref{eq:defCPT}, 
or more geometrically in Euclidean signature with gamma matrices $\gamma^\mu$,
a reflection along a direction $\hat{n}^\mu$ is given by $\hat{n} \cdot \gamma$ up to phase for fermions, and 
$(\hat{n} \cdot \gamma)(\hat{n}' \cdot \gamma) = -(\hat{n}' \cdot \gamma)( \hat{n} \cdot \gamma)$ if $\hat{n}$ and $\hat{n}'$ are orthogonal.
In this paper we will be concerned with the case $\sCP^2=(-1)^F$.

The $\sCP$ reverses all the spatial coordinates $\vec{x} \to -\vec{x}$. If we instead consider $\sCR$ \cite{Witten:2015aba} which only reverses one coordinate, 
say $x^1 \to -x^1$, we get 
\beq
\sCR^2=(-1)^F \sCP^2
\eeq in $3{+}1$ spacetime dimensions,
because of the difference by the $2\pi$ rotation on the $x^2 - x^3$ plane.
The cases $\sCR^2=1~(\leftrightarrow \sCP^2=(-1)^F)$ and $\sCR^2=(-1)^F~(\leftrightarrow \sCP^2=1)$ correspond to the $\Pin^+$  
and $\Pin^-$ symmetries, respectively, when the $\sCP$ is embedded in the double cover of the unoriented Lorentz group $\OO(3,1)$.
A review of these $\Pin$ symmetries can be found in e.g.~\cite{Kapustin:2014dxa,Witten:2015aba}.

\subsection{Single Majorana fermion}
Suppose that we have a single Weyl fermion $\psi_\a$ which has two components $\alpha=1,2$. 
The action of $\sCP$ is dictated by the Lorentz symmetry up to a phase,
\beq
\sCP( \psi_\alpha(t,\vec{x}) )=+ \ii \eta \bar{\psi}{}^\aa(t,-\vec{x}),  \label{eq:majCP}
\eeq
where we extracted a factor of $\ii$ from the phase, for later convenience.
By raising and lowering the indices, we get 
\beq
\sCP (\psi^\alpha(t,\vec{x}) )= - \ii \eta \bar{\psi}_\aa(t,-\vec{x}),  
\eeq
in the convention of Wess and Bagger\cite{WessBagger}, where the minus sign is due to the fact that $\epsilon^{\a\b}$ and $\epsilon_{\a\b}$ differ by a sign as a matrix. 
By taking complex conjugates, we also obtain
\beq
\sCP (\bar{\psi}_\aa(t,\vec{x}) ) &= -\ii \eta^* \psi^\a(t,-\vec{x}),  \\
\sCP (\bar{\psi}{}^\aa(t,\vec{x}) ) &=+\ii \eta^* \psi_\a(t,-\vec{x}).
\eeq
In particular, we have $\sCP^2 (\psi )=-\psi$.

Of course this minus sign can also be derived without any explicit computation. 
Note that $\sCP$ commutes with the spatial $\SU(2)$ rotation, and also that $\sCP$ sends the doublet $\psi$ of $\SU(2)$ to the complex conjugate $\bar{\psi}$. 
Such a transformation is possible only if the representation is strictly or pseudo real, 
and the transformation squares to $+ 1$ or $-1$ depending on whether it is strictly or pseudo real, respectively.
As is well-known, an irreducible half-integer spin representation of $\SU(2)$ is pseudo-real, meaning that $\sCP$ acting on an irreducible fermion needs to square to $-1$.

Now suppose that the fermion $\psi_\alpha$ has a real Majorana mass $m$,
\beq
 m(\psi^\a \psi_\a+\bar{\psi}_\aa \bar{\psi}{}^\aa).
\eeq
For the Majorana mass to respect the $\sCP$, we need
\beq
\bar{\psi}{}_\aa \bar{\psi}{}^\aa=\sCP  (\psi^\a \psi_\a ) =\eta^2 \bar{\psi}_\aa \bar{\psi}{}^\aa,
\eeq
from which we conclude $\eta=\pm 1$ .

Suppose we perform  a field redefinition 
\beq
\psi'=\ii\psi.
\eeq
This changes  the signs of both $m$ and $\eta$, but the combination
\beq
\eta \sign(m)
\eeq
is invariant. In this paper, when we speak of the sign of the mass term, we refer to this combination.

\subsection{Multiple Majorana fermions}\label{subsec:multi}
Let us extend $\sCP$ transformation to the case where there are multiple fermions $\psi_i$,  ($i=1,2,\ldots$).
We can define a $\sCP$ by
\beq
\sCP( (\psi_i)_\alpha(t,\vec{x}) )= \ii  {\eta}_{ij} \bar{(\psi_j)}^\aa(t,-\vec{x}),  
\eeq
where $ {\eta}$ is a unitary matrix and summation over $j$ is understood. 
The mass term is
\beq
 (\psi_i)^\a  {M}^{ij}(\psi_j)_\a+\bar{(\psi_i)}_\aa  ({M}^{ij})^* \bar{(\psi_i)}^\aa.
\eeq
where $ {M}^{ij}$ is a symmetric matrix $ {M}^T= {M}$, and the notation $*$ means taking complex conjugates of entries of the matrix.
By imposing the condition that this mass term is invariant under the $\sCP$, we get
\beq
 {\eta}^T  {M}  {\eta}= {M}^*, \label{eq:mmat}
\eeq
where we have used matrix notation.

By acting $\sCP$ twice, we get
\beq
\sCP^2 (\psi )=- {\eta}  {\eta}^* \psi
\eeq
where again we have used matrix notation. 
We would like to have $(\sCP)^2=(-1)^F$, or equivalently, 
we require that $\sCP$ is a part of the $\Pin^+(3,1)$ symmetry. 
Then $ {\eta}  {\eta}^* =1$.  Therefore
$\eta$ is a unitary symmetric matrix: $ {\eta} {\eta}^\dagger =1$ and $ {\eta}^T= {\eta}$.
Recalling \eqref{eq:mmat}, we have that
\beq
 {\eta}  {M} =( {\eta}  {M} )^\dagger,
\eeq
or equivalently,  $  {\eta} {M}$ is hermitian. 

Now let us perform a field redefinition
\beq
\psi = {V}\psi'
\eeq
by a unitary matrix $ {V}$. This effects the following changes to the various matrices introduced above:
\beq
 {M}'= {V}^T {M}  {V},~~~ {\eta}'=(V)^{-1}  {\eta} (V^T)^{-1},~~~ {\eta}' {M}'= {V}^{-1} {\eta} {M}  {V}.
\eeq
Therefore, the eigenvalues of $ {\eta} {M}$ are invariant under this redefinition. 
We can define an invariant $\nu( \eta{M})$ as the number of negative eigenvalues,
\beq
\nu( \eta{M})= \#(\text{negative eigenvalues of } {\eta M}).
\eeq
As we saw in Sec.~\ref{1.2}, this number classifies the free fermionic SPT phases.

\subsection{Gauge fields}
We are mainly interested in the case in which Majorana fermions are coupled to  a gauge group $G$ in some strictly real representation $\rho$.
The kinetic term of the fermions is
\beq
 -\ii \bar{\psi} \bar{\sigma}^\mu (\partial_\mu + \rho(T_A) A^A_\mu) \psi
\eeq
where $T_A~(A=1,\ldots,\dim G)$ are generators of the gauge group $G$ and $\rho(T_A)$ are the representation matrices of the generators, which are real antisymmetric matrices.
By  requiring  the $\sCP$ invariance of this kinetic term, we see that the $\sCP$ transformation acts on the gauge fields as follows:
\beq
\sCP  (A^A_0 )=A^A_0,~~~\sCP  (A^A_i) =-A^A_i~(i=1,2,3).
\eeq

\subsection{$\sCP$ transformation and supersymmetry}\label{subsec:susyCP}
Let us extend the $\sCP$ transformation to supersymmetric theories. First, supercharges must transform as
$
\sCP (Q_\alpha )= \ii \eta_Q \bar{Q}^\aa.
$
By a redefinition of the phase of $Q_\alpha$, we can always set $\eta_Q=1$. Then we get
\beq
\sCP (Q_\alpha )&= \ii  \bar{Q}^\aa,&
\sCP (Q^\alpha) &= -\ii  \bar{Q}_\aa,&
\sCP (\bar{Q}_\aa)&=- \ii  Q^\alpha,&
\sCP (\bar{Q}^\aa )&= \ii  Q_\alpha.
\eeq
Correspondingly, we define
\beq
\sCP(\theta_\alpha)&=-\ii\bar{\theta}^\aa,&
\sCP(\theta^\alpha)&=\ii\bar{\theta}_\aa,&
\sCP(\bar{\theta}_\aa)&=\ii\theta^\alpha,&
\sCP(\bar{\theta}^\aa)&=-\ii\theta_\alpha, \label{eq:thetaCP}
\eeq
so that we get $\theta^\alpha \sCP(Q_\alpha)=\sCP(\bar{\theta}_\aa)\bar{Q}^\aa$ and $\bar{\theta}_\aa \sCP(\bar{Q}^\aa)=\sCP(\theta^\a) Q_\alpha$.

Generally, for a superfield 
$K(\theta,\bar{\theta},x)$ whose lowest component is a scalar,
the rule of $\sCP$ transformation is
\beq
\sCP (K(\theta,\bar{\theta},x) )= \eta_K \bar{K}(\sCP(\theta),\sCP(\bar{\theta}), \sCP(x)) \label{eq:generalsuperCP}
\eeq
where $\eta_K$ is a phase factor, $\bar{K}(\theta,\bar{\theta},x)$ is the complex conjugate of $K(\theta,\bar{\theta},x)$, and $\sCP(t,x)=(t,-x)$.

This transformation law is as it should be if we admit the existence of the superspace, since once the action of a group on the (super)space is given, the action of the group on functions on that (super)space is naturally given by the pull-back. 
For those who prefer to start from just the super-algebra acting on the Hilbert spaces of the quantum theory,
the same transformation law \eqref{eq:generalsuperCP} can be derived also as follows. 
In that approach, general superfield $K(\theta,\bar{\theta},x)$, whose lowest component operator is $K(x)$, is {\it defined} by using the supercharges $Q$ and $\bar{Q}$ as
${K}(\theta,\bar{\theta},x)=e^{\ii (\theta Q +\bar{\theta} \bar{Q})} K(x)e^{-\ii (\theta Q +\bar{\theta} \bar{Q})}$. 
Then, the $\sCP$ transformation is given by 
\beq
\sCP({K}(\theta,\bar{\theta},x)) := e^{\ii (\theta \sCP(Q) +\bar{\theta} \sCP(\bar{Q}) )} \sCP(K(x)) e^{-\ii (\theta \sCP(Q) +\bar{\theta} \sCP(\bar{Q}) )}.
\eeq
Here we have used the fact that the meaning of $\sCP( \CO )$ acting on a quantum mechanical operator $\CO$ is  given by $(\sCP)\CO (\sCP)^\dagger$, with $(\sCP)$ now interpreted as the operator which acts on the Hilbert space. Then $(\sCP)$ in this operator sense commutes with $\theta$ and $\bar{\theta}$. We remark that
the precise meaning of ``$\sCP$" appearing in $\sCP(\theta)$ and $\sCP(\bar{\theta})$ is different from the one in the operator sense; 
$\sCP(\theta)$ and $\sCP(\bar{\theta})$ are just defined by \eqref{eq:thetaCP}.
We take the transformation of the lowest component as 
$\sCP(K(x)) = \eta_K \bar{K}(\sCP(x))$. Then, by using $\theta^\alpha \sCP(Q_\alpha)=\sCP(\bar{\theta}_\aa)\bar{Q}^\aa$ 
and $\bar{\theta}_\aa \sCP(\bar{Q}^\aa)=\sCP(\theta^\a) Q_\alpha$, we get \eqref{eq:generalsuperCP}.

As a special case, consider a chiral superfield 
\beq
\Phi=\phi+\sqrt{2}\theta\psi+\theta^2 F.
\eeq
If we require 
$
\sCP (\phi(t,\vec{x}))= \eta \bar{\phi}(t,-\vec{x}),  
$
then the above general rule determines
\beq
\sCP (\phi (t,\vec{x}) ) &=  \eta \bar{\phi}(t,-\vec{x}),  \\
\sCP (\psi_\alpha(t,\vec{x})) &= \ii \eta \bar{\psi}^\aa(t,-\vec{x}),   \\
\sCP (F (t,\vec{x}) )&=  \eta \bar{F}(t,-\vec{x}).
\eeq

Now let us discuss what action can preserve the $\sCP$ invariance. 
Notice that combinations such as $\theta^2=\theta^\alpha \theta_\alpha$ and $\bar{\theta}^2=\bar{\theta}_\aa \bar{\theta}^\aa$ satisfy \begin{equation}
\sCP(\theta^2)=\bar\theta^2,~~~\sCP(\bar{\theta}^2)=\theta^2.
\end{equation}
 This observation makes superspace analysis quite straightforward. 
First, the kinetic terms of chiral fields, including gauge superfields $V^A$, is given by
\beq
\int d\theta^2 d\bar{\theta}^2 \bar{\Phi} e^{2\ii \rho(T_A) V^A} \Phi
\eeq
where, as before, we consider a strictly real representation $\rho$, and $\rho(T_A)$ are anti-hermitian. 
It is easy to see that this kinetic term is invariant if the phase factor of $V$ is $\eta_V=1$,
i.e., 
\begin{equation}
\sCP (V^A(\theta,\bar{\theta},x) )=V^A(\sCP(\theta),\sCP(\bar{\theta}), \sCP(x)).
\end{equation}

Next, let us analyze the superpotential.
For simplicity, let us use a `Majorana basis' for the chiral fields $\Phi$ such that $\eta=+1$.
Consider a superpotential of the form
\beq
W = m_{ij}\Phi_i\Phi_j + y_{ijk} \Phi_i \Phi_j \Phi_k+ \cdots.
\eeq
The term in the action is then $\int d^2\theta W + \int d^2\bar\theta \bar W$.
The condition for the $\sCP$ invariance is simply that all the parameters $m_{ij}$, $y_{ijk}$, etc., are real. 
More general phase factor $\eta$ can be treated in the same way as in Sec.~\ref{subsec:multi}.

\section{$\sCP$ in concrete gauge theories}\label{sec:concreteCP}

Now let us discuss $\sCP$ invariance of some gauge theories. 
There is a remark on the terminology. What we call $\sCP$ is usually called  $\sP$ in standard textbooks in the case of $\SU(N)$ or $\Sp(N)$ gauge theories.
However, our interest is on the SPT phases related to free or interacting Majorana fermions, so our focus is centered around Majorana fermions rather than gauge fields.
Therefore we stick to call the symmetry $\sCP$ rather than $\sP$.

\subsection{Non-supersymmetric theories}
Here  we consider $G=\SO(N), \SU(N)$ or $\Sp(N)$ gauge theories with $N_f$ flavors of fermion fields in the defining representation.

\paragraph{$\SO(N)$ theories.}
The vector representation is  a strictly real representation,
so we can simply consider $N \times N_f$ Majorana fermions and then gauge them by $\SO(N)$.
By an appropriate redefinition of fields, the $\sCP$ transformation is just given by \eqref{eq:majCP} with $\eta=1$.
The Lagrangian is
\beq
 -\ii \bar{\psi} \bar{\sigma}^\mu D_\mu \psi -  \frac{1}{2} {m}_{ij} [\psi_i\psi_j+ \bar{\psi}_i\bar{\psi}_j],\label{eq:soLag}
\eeq
where $i,j,\ldots$ are flavor indices and $m_{ij}$ is a real symmetric matrix. 
The $D_\mu$ is the covariant derivative and gauge indices are suppressed.

In the notation of Sec.~\ref{subsec:multi}, we write
\beq
{M}=1 \otimes {m},~~~{\eta}=1 \otimes 1
\eeq
where in the notation $A \otimes B$, $A$ acts on gauge indices and $B$ acts on flavor indices.
Therefore, we get
\beq
\nu( \eta{M})=N \nu({m}).
\eeq

\paragraph{$\SU(N)$ theories.}
One flavor of quarks in the fundamental representation means that we introduce fermions in the representation ${\bf N}\oplus\bar{\bf N}$.
Each of ${\bf N}$ and $\bar{\bf N}$ is a complex representation, but their sum ${\bf N}\oplus\bar{\bf N}$ is a  strictly real representation and hence 
we can take the generators $\rho(T_a)$ to be real in a certain basis. Therefore, we can define $\sCP$ in the way discussed in Appendix~\ref{sec:CPTbasics}.
However, as it is sometimes  more convenient to use the usual complex basis,
let us define $\sCP$ directly there.

We introduce fermions $\psi^a_i$ and $\tilde{\psi}^i_a$, where $a=1,\ldots,N$ is a gauge index and $i=1,\ldots,N_f$ is a flavor index.
Notice that the $\sCP$ transformation and the $\SU(N) \times \SU(N_f)$ symmetry commute as can be seen in a Majorana basis.
We therefore define
\beq
\sCP ((\psi^a_i)_\alpha ) &= \ii ( \bar{\tilde{\psi}}{}^a_i )^\aa, &
\sCP ((\tilde{\psi}^i_a)_\alpha)  &= \ii ( \bar{\psi}{}^i_a )^\aa, 
\eeq
where we have used the notation that e.g., the complex conjugate of $ \psi^a_i$ is given by $\overline{(\psi^a_i) } = \bar{\psi}{}^i_a$
by exchanging the upper and lower indices, which makes the group action of $\SU(N) \times \SU(N_f)$ more transparent.

The Lagrangian is 
\beq
 -\ii \bar{\psi} \bar{\sigma}^\mu D_\mu \psi   - \ii \bar{\tilde{\psi}} \bar{\sigma}^\mu D_\mu \tilde{\psi}   
-  ({m}^i_{j} \tilde{\psi}^j_a\psi^a_i + ({m}^\dagger)^i_{j} \bar{\psi}{}^j_a  \bar{\tilde{\psi}}{}^a_i ). \label{eq:suLag}
\eeq
The $\sCP$ invariance requires that the matrix $m$ is hermitian: ${m}={m}^\dagger$.
This is just a special case of the more general discussion in Sec.~\ref{subsec:multi}. In fact, in the basis
\beq
(\psi^a_i, \tilde{\psi}^i_a)
\eeq
we have matrices
\beq
{M} &= \left( \begin{array}{cc}
0 &1 \otimes  {m}^T \\
1 \otimes {m} & 0 
\end{array}
\right),&
{\eta} &=  \left( \begin{array}{cc}
0 & 1 \otimes 1 \\
1 \otimes 1 & 0 
\end{array}
\right),
\eeq
Then, we get
\beq
 {\eta}  {M}=
 \left( \begin{array}{cc}
1 \otimes   {m} & 0 \\
 0 & 1 \otimes  {m}^T
\end{array}
\right).
\eeq
The requirement that the $ {\eta}  {M}$ is hermitian is equivalent to the requirement that $ {m}$ is hermitian.
We also get
\beq
\nu( \eta{M})=2N \nu( {m}).
\eeq

\paragraph{$\Sp(N)$ theories.}
The fundamental representation ${\bf 2N}$ of the $\Sp(N)$ is pseudo-real. If we have even number of copies of the fundamental representation,
the total representation can be made to be strictly real. So we take the number of fundamental Weyl fermions as $2N_f$; of course this is also required by 
cancellation of the global gauge anomaly.
Let us denote the fields as 
$\psi^a_i$, where $a=1,\ldots,2N$ and $i=1,\ldots,2N_f$.

The $\sCP$ commutes with $\Sp(N) \times \Sp(N_f)$ symmetry, and we define
\beq
\sCP ((\psi^a_i)_\alpha) = \ii  J^{ab} J_{ij}( \bar{\psi}{}^j_b )^\aa,
\eeq
where $J^{ab}$ and $J_{ij}$ are anti-symmetric invariant tensors of $\Sp(N)$ and $\Sp(N_f)$, respectively.
The Lagrangian is given by 
\beq
 -\ii \bar{\psi} \bar{\sigma}^\mu D_\mu \psi -  \frac{1}{2}[  {m}^{ij} (J^{-1})_{ab}\psi^a_i\psi^b_j -( {m}^*)_{ij} (J)^{ab} \bar{\psi}{}_a^i\bar{\psi}{}_b^j], \label{eq:spLag}
\eeq
where $ {m}$ is an anti-symmetric matrix $ {m}^T=- {m}$. The $\sCP$ invariance requires
\beq
 {m}^*=J  {m} J^T.\label{eq:spmass}
\eeq

In the notation of Sec.~\ref{subsec:multi}, we have
\beq
 {M}=J^{-1} \otimes  {m},~~~ {\eta}=J \otimes J,
\eeq
and hence
\beq
 {\eta} {M}=1 \otimes J {m}.
\eeq
The condition that $ {\eta} {M}$ is hermitian is equivalent to the condition \eqref{eq:spmass}. We get
\beq
\nu( {\eta M})=2N \nu(J {m}).
\eeq

\subsection{Supersymmetric theories}\label{subsec:susyLag}
It is straightforward to extend the  non-supersymmetric $\SO(N), \SU(N)$ or $\Sp(N)$ theories discussed above to \Nequals1  supersymmetric theories if we only consider
chiral fields in the fundamental representation.
Corresponding to the non-supersymmetric mass term
\beq
-\psi^i M_{ij} \psi^j,
\eeq
we just consider the superpotential
\beq
W= Q^i {M}_{ij} Q^j
\eeq
where $Q^i$ denotes quark superfields in the fundamental representation of the gauge group.
The $\sCP$ is extended as described in Sec.~\ref{subsec:susyCP}, and the condition for ${M}$ is completely the same as in the non-supersymmetric case.

When we extend the supersymmetry to \Nequals2, we have to introduce an adjoint chiral superfield $\Phi$.
Quarks also have to form hypermultiplets of \Nequals2  SUSY.
Consider hypermultiplets $(Q^a_i, \tilde{Q}^i_a)$ in a complex representation. Strictly real and pseudo real representations are just a special case of complex representation. 
The coupling of these hypermultiplets to the adjoint chiral field $\Phi$ is given by
\beq
W=\tilde{Q}^i_a  \rho(T_A)^a_b  \Phi^A Q^b_i.
\eeq
Here $\rho(T_A)$ is taken to be anti-hermitian.
Then we can consider the following $\sCP$ transformation
\beq
\sCP (Q^a_i ) &= \bar{\tilde Q}^a_i,&
\sCP (\tilde{Q}_a^i )&= \bar{ Q}_a^i,&
\sCP (\Phi^A ) &= - \bar{\Phi}^A (\leftrightarrow \sCP (\Phi ) =\Phi^\dagger)
\eeq
where we have suppressed superspace coordinates, and $\dagger$ acts on the matrix $\Phi=\Phi^A \rho(T_A)$ as hermitian conjugate.
We can also add mass terms which are consistent with the $\sCP$ as
\beq
W_\text{mass}=\tilde{Q}^i_a  {m}^j_i Q^a_j+\frac{1}{2} m_\Phi \Tr \Phi^2 ,
\eeq
where ${m}$ is hermitian. The first term is consistent with \Nequals2  SUSY, while the second term breaks \Nequals2  to \Nequals1 .

\section{The Wess-Zumino-Witten term}\label{sec:WZW}

\subsection{Definitions}
 Here we summarize the properties of the Wess-Zumino-Witten term. 
 It is not difficult to discuss it in arbitrary even dimensions $d=2n$.
For simplicity we assume in this subsection that manifolds are orientable.

Suppose that we have a theory with global symmetry $F_0$ which is spontaneously broken to $F \subset F_0$.
Also, suppose that $F_0$ has a 't~Hooft anomaly represented by the anomaly polynomial 
\beq
I_{2n+2} = 2\pi  \cdot \frac{\kappa}{(n+1)!} \tr \left( \frac{\ii {\cal F}}{2\pi} \right)^{n+1}
\eeq
where the trace is taken in some representation, $\kappa$ is the 't~Hooft anomaly coefficient, and 
${\cal F}=\frac{1}{2} {\cal F}_{\mu\nu} dx^\mu \wedge dx^\nu$ with ${\cal F}_{\mu\nu}=\partial_\mu \CA_\nu-\partial_\nu \CA_\mu + [\CA_\mu, \CA_\nu]$ 
is the field strength 2-form of the background gauge field of the flavor symmetry.
The $\kappa$ is defined such that the trace over all fermions under the symmetry $F_0$, denoted by $\Tr_\text{fermions}$, is given by $\Tr_\text{fermions} =\kappa \tr$.

The $I_{2n+2}$ is a $(2n+2)$-form, and we define $(2n+1)$-form $I_{2n+1}(\CA)$ as
\beq
d I_{2n+1}(\CA)=I_{2n+2}.
\eeq
Now, let $V $ be the Goldstone boson field which takes values in $F_0$. We impose the gauge invariance $V \sim VW$ for $W \in F$ so that $V$ 
is the variable taking values in $F_0/F$.
In this case, the WZW term (including the background field $\CA$) is given by
\beq
S_{\rm WZW}= \int_N (I_{2n+1} (\CA^V) -I_{2n+1}(\CA)) ~\text{up to manifestly local terms in $2n$-dim},
\eeq
where $N$ is an auxiliary $2n+1$ dimensional manifold whose boundary is the $2n$ dimensional manifold $M$, and 
\beq
\CA^V= V^{-1} \CA V+V^{-1} dV.
\eeq
One can check the following. (1) This action only depends on the boundary value of the fields modulo $2 \pi  \kappa $ because $\int I_{2n+1}(\CA)$ 
is the Chern-Simons action and we are taking the difference of the Chern-Simons actions $\int I_{2n+1} (\CA^V) -\int I_{2n+1}(\CA)$ which differ only by ``the gauge transformation by $V$". 
(2) This action reproduces the 't~Hooft anomaly under the $F_0$ flavor transformation
$\CA \to g \CA g^{-1}  +g dg^{-1}$ and $V \to gV$, because $\CA^V$ is invariant while the term $I_{2n+1}(\CA)$ gives the anomaly by the standard anomaly descent argument.
(3) Under the gauge transformation $V \to VW$ with $W \in F$, $\CA^V$ changes as $\CA^V \to W^{-1} \CA^V W+W^{-1} dW$ 
and hence gives anomaly from $I_{2n+1}(\CA^V)$
by the descent equation argument. This is zero (up to contributions which are cancelled by manifestly local counterterms in $2n$ dimensions) 
if the current of $F$ is free from 't~Hooft anomaly. Assuming that is the case, $S_{\rm WZW}$ is invariant under the transformation $W$ if we choose appropriate counterterms.
This assumption is satisfied in the theories considered in this paper.

More explicitly, $I_{2n+1}$ is given by
\beq
I_{2n+1}(\CA)=2\pi  \cdot \frac{\kappa}{n!} \int^1_0 dt \tr \left( \frac{ (t d{\cal  \CA}+t^2 \CA^2)^n\CA}{(-2\ii \pi)^{n+1}} \right) \label{eq:CS2n+1}
\eeq
and in particular,
\beq
I_{2n+1}(V^{-1} dV) &=2\pi  \kappa \cdot \frac{-1}{(2\pi \ii)^{n+1}} \frac{n!}{(2n+1)!} \tr (V^{-1}dV)^{2n+1} \nonumber \\
& :=2\pi  \kappa \cdot \Omega_{2n+1}(V).
\eeq

Instead of \eqref{eq:CS2n+1}, one can also use the following definition of $I_{2n+1}$ which differs from \eqref{eq:CS2n+1} by a total derivative.
(This paragraph is outside the main line of argument and may be skipped.)
Let us split the background gauge field as $\CA=\CA'+\CA''$, where $\CA'$ takes values in the Lie algebra of $F$, and $\CA''$ is orthogonal to the Lie algebra of $F$ inside $F_0$.
Then we have $I_{2n+2}(\CA)-I_{2n+2}(\CA')=d I'_{2n+1}(\CA',\CA'')$ where 
\beq
I'_{2n+1}(\CA',\CA'')=2\pi  \cdot \frac{\kappa}{n!} \int^1_0 dt \tr \left( \frac{ ({\cal F}'+t D'{\cal  \CA''}+t^2 \CA''^2)^n\CA''}{(-2\ii \pi)^{n+1}} \right).
\eeq
Here ${\cal F}'$ is the field strength of $\CA'$, and $D'$ is the covariant exterior derivative using $\CA'$.
If $F$ is free from 't~Hooft anomaly as assumed above, we get $I_{2n+2}(\CA')=0$ and hence $I_{2n+2}(\CA)=d I'_{2n+1}(\CA',\CA'')$ so we can use $I'_{2n+1}(\CA',\CA'')$
in the definition of WZW term. The point is that $I'_{2n+1}(\CA',\CA'')$ is manifestly invariant under the gauge transformation of $F$,
and hence the WZW term is manifestly invariant under $V \to VW$ without any counterterm.

\subsection{Computations using Clifford algebras}
Let us compute the integral of $\Omega_{2n+1}$ defined above for a few  specific configurations.\footnote{The argument here was reviewed in \cite{Witten:1998cd}.}
First, we take the gamma matrices in $2n+2$ dimensions $\Gamma_M$ which satisfy
\beq
\{ \Gamma_M, \Gamma_N\}=2\delta_{MN}.
\eeq
Their sizes are $2^{n+1} \times 2^{n+1}$. 
Let $\bar{\Gamma}=\ii^{n+1}\Gamma_1 \cdots \Gamma_{2n+2}$ the chirality matrix.

We consider a unit sphere $S^{2n+1}$ embedded in $\bR^{2n+2}$ with the coordinates $X^I$ $(I=1,\ldots,2n+2)$.
We denote the points on $S^{2n+1}$ by  $\hat{X}^M$ with $(\hat{X})^2=1$. 
We also take a specific point $\hat{X}_0^M$  on $S^{2n+1}$.
Then, we consider a configuration of the Goldstone field $V$ on $S^{2n+1}$ given by
\beq
V=P_{+}(\Gamma \cdot \hat{X}_0)(\Gamma \cdot \hat{X}),\label{eq:winding1}
\eeq
where $P_{+}=(1+\bar{\Gamma})/2$ is the chirality projection, $\Gamma \cdot \hat{X}=\Gamma_M \hat{X}^M$, and it should be understood that we only take
the block of the matrix which has positive chirality.
Then $V$ is a $2^n \times 2^n$ matrix and is unitary.

Let $\Gamma_{MN}=\frac{1}{2}(\Gamma_M \Gamma_N-\Gamma_M \Gamma_N)$. Then
one can check that $V^{-1}dV=P_{+} \Gamma_{MN}\hat{X}^M d\hat{X}^N$ and 
$(V^{-1}dV)^2=-d(V^{-1}dV)=-P_{+} \Gamma_{MN}d\hat{X}^M d\hat{X}^N$. By using these equations we get by a straightforward computation that
\beq
\int_{S^{2n+1}} \Omega_{2n+1} =\int_{D^{2n+2}} d\Omega_{2n+1} 
=1,\label{eq:topinv}
\eeq
where $D^{2n+2}$ is the disk bounded by $S^{2n+1}$, and we have extended $\Omega_{2n+1}$ to $D^{2n+2}$ 
by replacing $\hat{X} \to X$ for $X \in D^{2n+2}$, i.e. \begin{equation}
d\Omega_{2n+1} = \frac{1}{(-2\pi \ii)^{n+1}} \frac{n!}{(2n+1)!}  \tr P_{+} (\Gamma_{MN}d{X}^M d{X}^N)^{n+1}.
\end{equation}
The configuration considered above is known to give the 
smallest absolute value of $\int \Omega_{2n+1} $ on a closed manifold.

Finally, notice the following property of $V$ as a function of $\hat{X}$. If we take the hermitian conjugate, we get
\beq
V(\hat{X})^\dagger=V( -\hat{X}+2\hat{X}_0 (\hat{X}_0 \cdot \hat{X}) ).\label{eq:hconjugate}
\eeq
This means that the coordinates of the directions orthogonal to $\hat{X}_0$ flip sign by the hermitian conjugate.
We will use this property in Sec.~\ref{sec:phasefromgoldstone}.

\section{More details on S-duality}\label{sec:Sd}
In this appendix we give more detailed description of the continuous deformation from small electric gauge coupling region $g \ll 1$ to large coupling region $g \gg 1$ which
corresponds to small magnetic gauge coupling $\dualg \ll 1$. For this purpose, it is most convenient to use the class S description which is manifestly symmetric under S-dual.
See e.g., \cite{Tachikawa:2013kta} and references therein for the background of this Appendix.

We consider $A_1=\SU(2)$ class S theory on a Riemann sphere with four regular punctures which corresponds to the $\CN=2$ $\SU(2)$ theory with $N_f=4$ flavors.
The $\SO(8)$ flavor symmetry has the subgroup \eqref{eq:foursu2}
\beq
\SU(2)_1 \times \SU(2)_2 \times \SU(2)_3 \times \SU(2)_4   \subset \SO(4) \times \SO(4) \subset \SO(8) ,
\eeq
and each puncture which we denote $p_i~(i=1,2,3,4)$ is associated to $\SU(2)_i$.

For simplicity, we consider mass parameters $(m_a,m_a,m_b,m_b)$ for the four flavors quarks.
This corresponds to the case where the punctures $p_1$ and $p_3$ have the mass parameters $m_a$ and $m_b$ respectively,
while the punctures $p_2$ and $p_4$ do not have any mass parameters.

Let $z$ be the coordinate of the Riemann sphere regarded as $\bC \cup \{ \infty \}$.
The $\sPinv$ comes from a Lorentz transformation of the 6d $\CN=(2,0)$ theory, and hence it acts on $z$ as an orientation reversing anti-holomorphic automorphism.
Its action can be taken as $z \to \bar{z}$. The positions of the punctures should also be fixed by this action.\footnote{There is another logical possibility that
two punctures are exchanged under $\sPinv$, but in that case $\sPinv$ does not commute with $\SO(8)$.}
So the punctures must be aligned on the $S^1$ given by $z=\bar{z}$. We take them to be
\beq
p_1: z=\infty,~~~p_2:z=1,~~~p_3:z=0, ~~~p_4:z=q.
\eeq
where $q \in \bR$ is a real parameter corresponding to the gauge coupling in a certain way.
There are three possible regions $0<q<1$, $q<0$ and $q>1$, and we will argue that our setup corresponds to $0<q<1$.
This fixes the cyclic order of the punctures on $S^1$ as $p_1,p_2,p_4,p_3$.

The Seiberg-Witten curve of the system is given by
\beq
\lambda^2= \frac{ m_a^2 z^2-uz+ qm_b^2  }{z^2(z-q)(z-1)} dz^2
\eeq
where $\lambda$ is the Seiberg-Witten differential. This is determined by requiring that $\lambda$ has poles with residues $m_a$ and $m_b$ at $p_1~(z=\infty)$ and $p_3~(z=0)$,
respectively.
From the above $\sPinv$ action, we can easily see that the $u$ transforms under $\sPinv$ (and hence $\sCP$)
as $\sPinv(u)=\bar{u}$.

The singular points on the $u$-plane can be found as the positions where the curve degenerates. 
The possibilities are either that the polynomial $m_a^2 z^2-uz+ qm_b^2 $ (i) has a degenerate root, or (ii) has a zero at $z=1$ or $z=q$.
Then we find singular points as
\beq
\text{A}:u=-2\sqrt{q}m_am_b,~~~\text{B}:u=+2\sqrt{q}m_am_b,~~~\text{C}:u=m_a^2+qm_b^2,~~~\text{C}':u=qm_a^2+m_b^2.
\eeq
If $q$ is a small positive value $0<q \ll 1$, these positions are precisely as expected in the field theory shown in the left hand side of Fig.~\ref{fig}.
Another possible region of $q$, given by $q<0$ do not reproduce our expectation. 
This is because if $q$ is negative, the points A and B are pure imaginary and they are exchanged under $\sCP$. 
This case corresponds to the case with the theta angle given by $\theta=\pi$. The region $q>1$ is just equivalent to $0<q<1$ by 
reparametrizations $z \to z^{-1}$, $q \to q^{-1}$, $m_a \leftrightarrow m_b$ and $u \to q^{-1}u$.
Therefore, we can focus our attention to $0<q<1$, and $q \ll 1$ corresponds to the small electric coupling $g \ll1$.
Then the region $0 <1-q \ll 1$ should correspond to the large electric coupling or equivalently small magnetic coupling.
Indeed, if we take $m_a=m_b=m$, and renormalize $u$ as $u'=u-\frac{1}{2}(1+\sqrt{q})^2m^2$, the singular points are located as
\beq
\text{A}:u'=-(2m)^2+O(1-\sqrt{q}),~~~\text{B}: u'=-\frac{(1-\sqrt{q})^2}{2}m^2,~~~\text{C}:u'=+\frac{(1-\sqrt{q})^2}{2}m^2,
\eeq
which reproduce the situation in the right hand side of Fig.~\ref{fig}.
In summary, just by changing $q$ in the region $0<q<1$ from $q \sim 0$ to $q \sim 1$, 
we can smoothly go from weakly coupled electric description to weakly coupled magnetic description.

Finally, let us comment on what the S-duality is. The above analysis using the Riemann sphere is manifestly symmetric under the S-duality.
However, we can make a change of the coordinate of the Riemann sphere from $z$ to $z'=1-z$. This exchanges the positions of the punctures $p_2$ and $p_3$
and also change the parameter $q$ to $q'=1-q$. This coordinate change corresponds to the S-duality.
Of course the physics is independent of the coordinate system, and hence the theory has the S-duality.
Under the S-duality which exchanges $p_2$ and $p_3$, the symmetry groups $\SU(2)_2$ and $\SU(2)_3$ are exchanged.
This is the fact used in Sec.~\ref{subsec:dualCP}.

\section{Explicit analysis of the fermion mass matrix on the dual side}\label{sec:mass}
Here, we analyze the  mass matrix of the fermions explicitly, as a complement to Sec.~\ref{subsec:continuous}.
In this appendix we use an abbreviation that the scalar components of the quarks $\dualQ_1$ and $\tilde{\dualQ}^1$ are just denoted as $\dualQ$ and $\tilde{\dualQ}$
and their fermionic components are denoted as $\frakq$ and $\tilde{\frakq}$. The $\CN=1$ gauginos are denoted as $\lambda$ and the fermions in $\dualPhi$ are denoted as $\psi$.

The mass terms involving gauginos $\lambda$ are given by
\beq
\dualQ^\dagger \lambda \frakq+ \bar{\tilde{\dualQ}} (- \lambda^T) \tilde{\frakq}^T + \frac{1}{\dualg^2}\Tr \dualPhi [\lambda, \psi],
\eeq
where we take
\beq
&\dualQ=(z, \bar{w})^T, ~~\tilde{\dualQ}=(\bar{z}, -w), ~~\frakq=(\frakq^+,\frakq^-)^T,~~\tilde{\frakq}=(\tilde{\frakq}_+, \tilde{\frakq}_-), \\
&\psi=\left(\begin{array}{cc} 
\psi^0 & \psi^{++} \\
\psi^{--} & -\psi^0
\end{array} \right),~~
\lambda=\left(\begin{array}{cc} 
\lambda^0 & \lambda^{++} \\
\lambda^{--} & -\lambda^0
\end{array} \right).
\eeq
Notice that $\tilde{\frakq}_+$ and $\tilde{\frakq}_-$ have the dual $\dualU(1)$ charges $-1$ and $+1$ respectively; 
one might want to write them as $\tilde{\frakq}_\pm :=\tilde{\frakq}^\mp$.
The other part of mass terms are determined by the superpotential. By a straightforward computation, we get the full mass matrix as
\beq
\Lambda^T M \Lambda
\eeq
where
\beq
\Lambda=(\lambda^{++},\psi^{++},\frakq^+,\tilde{\frakq}_-;\lambda^{--},\psi^{--},\tilde{\frakq}_+, \frakq^-;  \lambda^0,  \psi^0)^T
\eeq
and the matrix $M$ is given by
\begin{equation}
M=\left(
\begin{array}{c|c|c}
O& A & X \\
\hline
A^T & O & Y \\
\hline
X^T & Y^T & Z
\end{array}
\right)
\end{equation}
where 
\begin{gather}
A=\begin{pmatrix}
0&4m_0/\dualg^2&\bar{w}&\bar{z}\\
   -4m_0/\dualg^2&m_\dualPhi&-\bar{z}&-\bar{w}\\
   w&w&2(m-m_0) & 0\\
  -z&-z&0&2(m+m_0)
\end{pmatrix},\\
X=\begin{pmatrix}
0&0 \\
0&0 \\
\bar{z}& -\bar{z}\\ 
-\bar{w}&\bar{w} 
\end{pmatrix},\quad
Y=\begin{pmatrix}
0&0 \\
0&0 \\
-{z}& -{z}\\ 
-{w}&-{w} 
\end{pmatrix},\quad
Z=\begin{pmatrix}
0 & 0\\
0& 2m_\dualPhi
\end{pmatrix}.
\end{gather}

It is not too hard to analyze this matrix analytically. First recall that when $m \neq 0$, only one of the $z$ or $w$ gets a vev.
Then, the above mass matrix has a $\U(1)$ symmetry which comes from the diagonal part of $\dualU(1) \times \U(1)_{\rm flavor}$ which is unbroken by the vev.
For example, let us consider the case $z=0$ and $w \neq 0$. (The other case is completely analogous.) 
Due to the $\U(1)$ symmetry, the fermions splits to
\beq
\Lambda_1=(\lambda^{++},\psi^{++},\frakq^+)^T,~~\Lambda_2=(\lambda^{--},\psi^{--},\tilde{\frakq}_+)^T,~~\Lambda_3=(\tilde{\frakq}_-, \frakq^-,  \lambda^0,  \psi^0)^T,
\eeq
and
\beq
\Lambda^T M \Lambda=(\Lambda_1^T M' \Lambda_2+\Lambda_2^T M'^T \Lambda_1) + \Lambda_3^T M'' \Lambda_3,
\eeq
where $M'$ and $M''$ can be read off from the above matrix.
By a straightforward computation, one gets
\beq
\det M' & = -2(m_0-m) \left( \frac{4m_0}{\dualg^2} \right)^2-|w|^2 \left( m_\Psi +\frac{8m_0}{\dualg^2} \right), \nonumber \\
\det M'' &= 4|w|^2 (2m_\dualPhi(m_0+m)+|w|^2).
\eeq
From these equations we can see $\det M' <0$ and $\det M'' >0$, so both of them are nonzero. 

When $m=0$, we can argue in the following way. For simplicity we consider the parameter region where $0<\dualg ^2 m_\dualPhi - 2m_0 \ll m_0$.
Then the vevs of $z$ and $w$ are much smaller than the other mass parameters. 
If we set $z=w=0$, the mass matrix $M$ is quite simple and one can see that all the fermions other than $\lambda^0$ are massive.
Then, we treat $z$ and $w$ as a perturbation. The $\lambda^0$ gets a mass from the mixing with $(\frakq^+,\frakq^-, \tilde{\frakq}_+,\tilde{\frakq}_-)$.
After integrating out massive fermions $(\frakq^+,\frakq^-, \tilde{\frakq}_+,\tilde{\frakq}_-)$, one finds that the mass term of $\lambda^0$ is given as
\beq
-\frac{|z|^2+|w|^2}{2m_0} \lambda^0 \lambda^0.
\eeq
This establishes that there are no massless fermions when we interpolate $(z \neq 0,w=0)$ and $(z =0, w \neq 0$),
since $|z|^2+|w|^2= (4\dualg ^2m_\dualPhi m_0 - 8m_0^2)/\dualg^2  \neq 0$ during this interpolation.

\bibliographystyle{ytphys}
\baselineskip=.9\baselineskip
\bibliography{ref}

\end{document}